\documentclass[sigconf,screen,openany,nonacm]{acmart}

\usepackage{graphicx}
\usepackage{tabularx}
\usepackage{booktabs} 
\usepackage{subfig} 
\usepackage{float}
\usepackage{color}
\usepackage{tikz}
\usetikzlibrary{shapes,arrows,patterns,positioning} 
\usepackage{caption}
\usepackage[normalem]{ulem}

\usepackage{longtable}
\usepackage{lipsum}
\usepackage{tabularx}
\usepackage{booktabs} 

\usepackage{multirow}
\usepackage[ruled,linesnumbered]{algorithm2e}
\usepackage{amsmath}
\usepackage{empheq}

\usepackage{ulem}

\usepackage{xspace,tcolorbox}
\usepackage{hyperref} 
\usepackage[utf8]{inputenc}
\usepackage{textcomp}
\usepackage{{enumitem}} 

\usepackage[symbol]{footmisc}

\usepackage[flushleft]{threeparttable}

\tikzstyle{decision} = [diamond, draw, 
    text width=4.5em, text badly centered, node distance=3cm, inner sep=0pt]
\tikzstyle{block} = [rectangle, draw, 
    text width=5em, text centered, rounded corners, minimum height=4em]
\tikzstyle{line} = [draw, -latex']
\tikzstyle{cloud} = [draw, ellipse,fill=red!20, node distance=3cm,
    minimum height=2em]

\newcommand{\revise}[1]{{\color{blue}#1}}

\usepackage{xargs} 
\usepackage[colorinlistoftodos,prependcaption,textsize=tiny]{todonotes}
\newcommandx{\commentt}[2][1=]{\todo[linecolor=red,backgroundcolor=red!25,bordercolor=red,#1]{#2}}

\usepackage[formats]{listings}
\usepackage{xcolor}

\AtBeginDocument{%
  }

\setcopyright{acmlicensed}
\copyrightyear{2024}
\acmYear{2024}
\acmDOI{XXXXXXX.XXXXXXX}

\begin{document}
\sloppy



\title{Excavating Vulnerabilities Lurking in Multi-Factor Authentication Protocols: A Systematic Security Analysis}

\thispagestyle{plain}
\pagestyle{plain}

\author{Ang Kok Wee}
\authornote{Both authors contributed equally to this work and they share the first authorship.}
\affiliation{%
  \institution{Singapore University of Technology and Design}
  \city{Singapore}
  \country{Singapore}
  }
\email{kok\_ang@alumni.sutd.edu.sg}

\author{Eyasu Getahun Chekole}
\authornotemark[1]
\affiliation{%
  \institution{Singapore University of Technology and Design}
  \city{Singapore}
  \country{Singapore}
  }
\email{eyasu\_chekole@sutd.edu.sg}

\author{Jianying Zhou}
\affiliation{%
  \institution{Singapore University of Technology and Design}
  \city{Singapore}
  \country{Singapore}
  }
\email{jianying\_zhou@sutd.edu.sg}

\begin{abstract}
Nowadays, cyberattacks are growing exponentially, 
causing havoc to 
Internet users. In particular, authentication attacks constitute the major attack vector 
where intruders impersonate legitimate users to maliciously access systems or resources. 
Traditional single-factor authentication (SFA) protocols are often bypassed by side-channel and other attack techniques, hence they are no longer sufficient to 
effectively address 
the current authentication requirements. 
To alleviate this problem, multi-factor authentication (MFA) protocols have been 
widely adopted recently, which helps to raise the security bar against imposters. 
Although MFA is generally considered more robust and secure than 
SFA, it may not always guarantee enhanced security and efficiency. This is because, critical security vulnerabilities and performance problems may still arise 
due to 
design or implementation flaws of the protocols. Such vulnerabilities are often left unnoticed by the designers or users until they are exploited by attackers. Therefore, the main objective of this work is identifying 
such vulnerabilities in existing MFA protocols by systematically analysing their designs and constructions. 
To this end, we first form a set of security evaluation criteria, encompassing both existing and newly introduced ones, 
which we believe are very critical for the security of MFA protocols. Then, we 
thoroughly review several MFA protocols across different domains. Subsequently, we revisit and 
thoroughly analyze the design and construction of the protocols to identify potential 
vulnerabilities. 
Consequently, we manage to identify critical vulnerabilities in ten of the MFA protocols investigated. 
We thoroughly discuss the identified vulnerabilities in each protocol and 
devise relevant mitigation strategies 
for each of the vulnerabilities identified. We also consolidate the performance information of the protocols. We believe that the consolidated security analysis and performance information 
would serve as a single reference point for researchers and practitioners to be aware of the potential security and performance issues when designing MFA protocols. 
This investigation also reinforces the fundamental need for an enhanced and 
secure design and implementation 
of MFA protocols. 
\end{abstract}

\maketitle



\section{Introduction}\label{introduction}

The importance of information security is increasing exponentially due to the dynamic cyber threats posed by malicious adversaries\cite{altulaihan2022cybersecurity,wong2023security}. They typically exploit various types of security weaknesses and flaws 
to achieve their malicious intent\cite{humayun2020cyber,ang2023vulngen}. In particular, authentication-related attacks constitute 
the major cyber-attacks in the cybersecurity landscape. 
In this regard, attackers exploit weaknesses in authentication protocols and impersonate legitimate users to gain unauthorized access to a system or service~\cite{impersonation_attack}. 
Therefore, employing effective 
authentication mechanisms is crucial to 
alleviate the prevalence of cyber risks in various domains. 

Traditionally, authentication protocols are based on a single authentication factor, e.g., passwords, PINs, preshared keys, and biometrics. 
However, such authentication protocols are no longer 
sufficient 
due to various reasons. Weak passwords and password-related vulnerabilities are prevalent concerns. Users frequently choose easily guessable passwords or reuse them across multiple accounts, providing opportunities for attackers to crack passwords and gain unauthorized access \cite{inbook_2}. Phishing attacks are also other significant threats in which attackers exploit human vulnerabilities by using fraudulent emails or fake login pages to trick users into revealing their authentication credentials. These attacks can be phenomenally successful even against users with strong passwords \cite{article_3}. Various side-channel techniques, brute-force attacks, or a combination of the two could allow attackers to leak passwords/PINs or preshared keys. Adversaries can also maliciously copy biometric features or hack into a victim's database in numerous ways. The need for more robust security measures has led to the emergence of multi-factor authentication (MFA) \cite{article_4}.  

MFA is often considered to be more robust and secure than SFA. 
However, it is crucial to recognize that 
only enforcing MFA may not always guarantee 
enhanced security. This is because, 
security weakness and flaws (vulnerabilities) may still arise due to an incorrect design or implementation of the MFA protocols. Since the construction of MFA protocols involves different authentication factors, various secret and public parameters and complex technical issues, 
it remains challenging to properly detect 
security flaws in such protocols. In fact, even rigorous formal security proofs usually fail to detect such flaws~\cite{article_50}, which is also proven in our analysis discussed below. Consequently, such vulnerabilities are often left unnoticed by the protocol designers or end-users until they are 
exploited by attackers. If exploited, they may result in disastrous impacts on users. 
Furthermore, some MFA protocols employ three or more authentication factors to enhance their security guarantee \cite{article_38,article_27,article_45,article_26}. However, 
increasing the number of factors or using heavyweight approaches may also impose heavy performance penalties, which might not be tolerable in certain resource-constrained systems. 

Therefore, this work aims to detect potential 
vulnerabilities in existing MFA protocols by systematically 
analyzing its 
design and construction information that is publicly available in the respective paper. 
To this end, we first perform a background study 
on MFA protocols. Specifically, we first discuss about authentication factors, which are the building blocks of 
MFA protocols. The design and construction of MFA protocols 
 involves a systematic synergy of different authentication factors. In fact, the security and robustness of MFA protocols is heavily 
 dependent on the synergy and the types of authentication factors used. 
 To better comprehend our study and analysis, we revamp the taxonomy of authentication factors in MFA (cf. Figure \ref{fig:image1}). In brief, we classify 
them into two main categories: \emph{conventional} and \emph{emerging} authentication factors. The former involves the commonly used authentication factors and are further classified as \emph{knowledge factors} (e.g., passwords, PINs and security questions), \emph{possession factors} (e.g., physical tokens, security keys, smart cards and mobile authenticator apps), \emph{inherent factors}~\cite{inproceedings_12} (e.g., fingerprints\cite{fingerprint_auth}, facial recognition\cite{solomon2022uface,solomon2023self}, iris scans\cite{iris_auth}, and image recognition\cite{solomon2023nearest,solomon2023face2}) and \emph{location factors}~\cite{article_13,gps_location_factor,ambient_location_factor,ip_geolocation_factor}. The latter includes recently introduced authentication factors that 
show high effectiveness, especially in the context of machine-to-machine (M2M) authentications. These includes \emph{historical data}~\cite{inproceedings_16,article_38}, \emph{physically unclonable functions (PUF)}~\cite{article_14,inproceedings_80,} and \emph{firmware integrity}~\cite{article_15,chekole2020cima,chekole2021scope}. Such a systematic classification of the authentication factors is imperative as the use of distinct and independent authentication factors plays a crucial role in improving the security and robustness of MFA protocols. The introduction of various machine-learning techniques 
\cite{solomon2023fass,solomon2023hdlhc,solomon2023face,solomon2024federated} also enhanced the seamless integration and robustness of different authentication factors in MFA.    
 
Then, we thoroughly conduct a literature review on 
various MFA protocols. To be more comprehensive, we cover several relevant protocols in different domains, 
which we classify them as generic client-server systems~\cite{article_55,client_server_new1,client_server_new2}, cloud computing~\cite{article_18,article_19,article_20}, finance~\cite{inproceedings_21,article_22,article_23,mfa_ml1}, healthcare~\cite{article_24,article_25,article_26,article_57}, generic IoT~\cite{inproceedings_49,article_15,article_27,article_28,article_56}, healthcare IoT~\cite{article_29,article_30,article_31,article_32,mfa_iomt}, industrial IoT (IIoT)~\cite{article_33, article_34, article_35, article_36, inbook_37, article_38, article_39,mfa_iiot_new1}, smart cities/home~\cite{article_40,article_41,article_42,article_43}, and wireless sensor networks (WSN)~\cite{article_44,article_45}. We also highlight the key security requirements and constraints under each domain. We also emphasise on IoT-based multi-factor authenticated key exchange (MAKE) protocols as they involve more critical security and efficiency requirements.  

First, we identify several protocols based on its relevance and recency. Then, we revisit and 
thoroughly analyze the design and implementation of the MFA protocols to identify potential vulnerabilities. 
More specifically, we delve into the intricate details of their constructions to identify 
vulnerabilities that can potentially jeopardize the security of the authentication process and future key secrecy. 
The analysis is performed heuristically 
based on a set of evaluation criteria we employed for this purpose. 
While different users adopt different set of security evaluation criteria, we formed our own set of criteria. It encompasses both existing and our newly introduced ones, 
which we believe are very critical for the security of MFA protocols. Then, we thoroughly evaluate several MFA protocols based on the formed criteria. Consequently, we manage to 
detect significant vulnerabilities in ten of the MFA protocols investigated, which could be readily exploited by an attacker. 
We thoroughly discuss the identified vulnerabilities and consolidate the performance information of the protocols. 

The identified vulnerabilities are related to the lack of: explicit mutual authentication, independence of authentication factors, distinctiveness of authentication factors, 
leakage resilience, perfect-forward secrecy, user anonymity, resilience against known attacks, and realistic adversarial assumptions. Finally, we propose relevant mitigation strategies for the identified vulnerabilities. We believe that the consolidated information provided would serve as a single reference point for researchers and practitioners to be aware of the potential security issues when designing MFA protocols. It also helps to apply the necessary mitigation strategies to the vulnerable ones. 



We believe that this work can provide valuable insights 
to security researchers and practitioners to better understand potential vulnerabilities that may 
exist in the design of MFA protocols and the attack vectors that could be utilized by adversaries. 
Remarkably, most of these vulnerable protocols we identified are published in top cybersecurity journals and conferences. However, those vulnerabilities went unnoticed during the design, implementation, testing, and peer-review processes. Although most of the authors provided rigorous formal proofs on their protocols, they failed to detect those flaws. This 
implies that heuristic analysis could sometimes be even more effective than formal proofs in certain contexts. 
Therefore, we believe that insights from this analysis could serve as the basis for proposing effective mitigation strategies to enhance the design of MFA protocols per the established security evaluation criteria. Outcomes of this work can also significantly contribute to the ongoing efforts in the research community to strengthen authentication protocols and mitigate the security risks of improper MFA designs. 

While there are existing works that tried to detect vulnerabilities in MFA protocols, they only cover certain security issues on some 
protocols using different evaluation criteria~\cite{article_44,inproceedings_11,inproceedings_53,inbook_2}. Some others provide a survey on the security of MFA protocols~\cite{article_47,inproceedings_46,article_4,article_44,article_9}, but they did not perform any 
new security analysis to detect vulnerabilities in the design and construction of the MFA protocols. To the best of our knowledge, 
this work is the first to provide a comprehensive security analysis, especially on design- and construction-level vulnerabilities of MFA protocols, using a new set of security evaluation criteria. 

Overall, the main objectives of this work are: 1) 
detecting critical security flaws in MFA protocols and report them before they are exploited; 2) showing that simply adding multiple authentication factors may not always guarantee enhanced security; 3) testifying the relevance of a heuristic analysis in such contexts (even performs better than formal analysis in some cases); 4) creating awareness on the critical design flaws in MFA protocols and provide mitigation strategies; and 5) providing insights and mitigation strategies towards the design of more secure MFA schemes. 

In sum, this work makes the following main contributions.
\begin{itemize}
    \item We perform a systematic review of several MFA protocols across different domains, which can serve as a single-point of reference about the state-of-the-art MFA protocols. 
    \item Our work goes beyond the conventional survey work. Because, 
    we also systematically analyze the security of several MFA protocols. To this end, 
    \begin{enumerate}
        \item We first form a set of security evaluation criteria (by introducing new ones and adopting some existing ones) that can be used to critically assess the security of MFA protocols.
        \item We thoroughly evaluate the protocols based on the formed criteria and managed to identify several critical vulnerabilities in ten of the protocols.
    \end{enumerate}
    \item We devise appropriate mitigation strategies for the vulnerabilities identified, based on our own perspectives and from certain existing sources. 
    \item We believe that this work provides sufficient 
    insights to the community about design-level security weaknesses and flaws in MFA protocols. 
\end{itemize}





\emph{Organization.} The remainder of this article is structured as follows. Section \ref{background} provides the relevant background on MFA and authentication factors. Section \ref{literature_review} provides our reviews of existing MFA protocols across various domains. Section \ref{threat_model} 
highlights our adversarial assumptions used to assess the security of MFA protocols. In Section \ref{security_analysis}, we thoroughly analyze and discuss the flaws and vulnerabilities of the selected MFA protocols. In Section \ref{mitigation_strategies}, we devise relevant mitigation strategies for the vulnerabilities identified. Section \ref{conclusion} concludes the article by outlining relevant future works. 
\section{Background}\label{background}

\subsection{Overview of MFA}
MFA is employed in diverse contexts to ensure secure access to systems, applications, or sensitive information. It strengthens security measures by requiring individuals or entities to provide multiple authentication factors that prove their identity before they are allowed to access the system or application. This approach goes beyond traditional SFA and adds additional layers of security by combining multiple authentication factors. 

As highlighted in the introduction, the reliance on a single authentication factor, such as passwords, PINs, secret keys, and biometrics, is no longer sufficient for the current security trends and requirements. So, by reducing the dependence on a single authentication factor, MFA can impose stronger verification mechanisms. Even if one of the factors is compromised, the inclusion of additional factors in MFA prevents attackers from advancing without presenting the complete set of authentication factors. It is worth noting that MFA is nowadays commonly required in regulatory standards \cite{misc_5, misc_6} and recommended in advisories \cite{misc_7, misc_8} to safeguard the security of systems and sensitive information. 

\begin{figure*}[htp]
    \centering
    \includegraphics[scale=0.27]{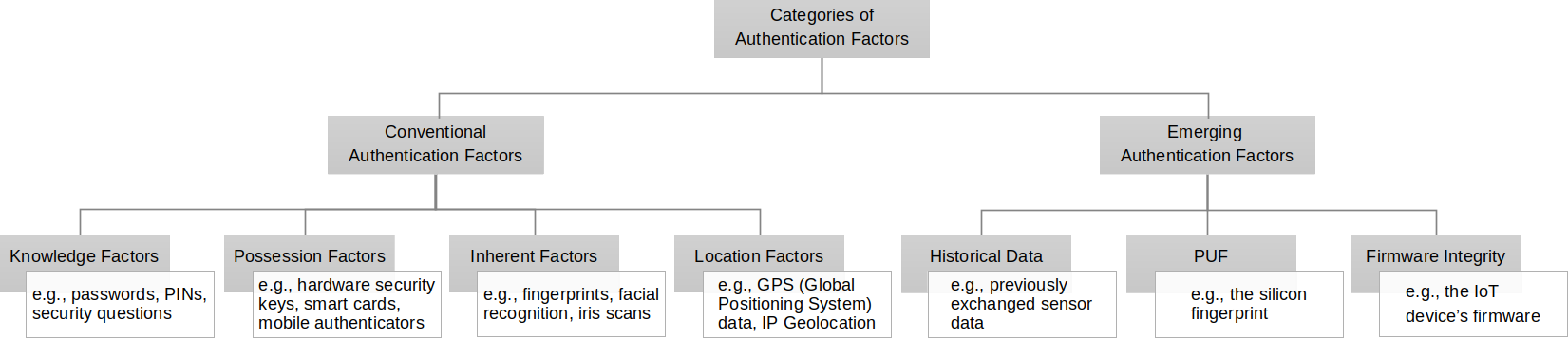}
    \caption{Taxonomy of authentication factors}
    \label{fig:image1}
\end{figure*}

Authentication factors are the building blocks of MFA protocols. They are combined systematically to form an MFA protocol with adequate security guarantees. 
In this work, we classify them into conventional and emerging categories. These categories and their respective subcategories are discussed in detail below and illustrated in Figure \ref{fig:image1} with examples.

\subsection{Conventional authentication factors}

The four categories of authentication factors typically used in MFA protocols are knowledge factors, possession factors, inherent factors, and location factors \cite{article_9}. 

\subsubsection{Knowledge factors}

Knowledge factors refer to something the individual knows, such as a password, PIN, or security questions. It is widely employed to verify an individual's identity and grant access to systems. 
Knowledge factors are based on the assumption that the individual is solely the one who knows the information. However, knowledge factors may potentially be susceptible to known attacks. As highlighted in earlier, passwords can be weak or easily guessed. Security questions can sometimes have answers that can be easily obtained or guessed \cite{inproceedings_10}. 

\subsubsection{Possession factors}

Possession factors involve something the individual possesses. A typical example of a possession factor is a physical token, such as a hardware security key or a smart card. 
Using such tokens, cryptographic algorithms are typically utilised to generate a unique code or response to verify the user's identity. Mobile authenticators are another form of a possession factor. They are applications installed on an individual's mobile device. Mobile authenticators employ algorithms such as HMAC-based One-Time Password (HOTP) or Time-based One-Time Password (TOTP) to generate a One-Time Password (OTP) at regular intervals. The individual is required to enter the generated OTP to verify their identity. 

\subsubsection{Inherent factors}

Inherent factors rely on something the individual is or has, typically related to an individual's biological traits or physical characteristics. These traits are difficult to replicate and provide an elevated level of security \cite{inproceedings_12}. Fingerprints, facial recognition, and iris scans 
are examples of inherent factors. 

 These inherent factors offer several advantages in authentication. They provide high accuracy and security since they are difficult to counterfeit or manipulate. Additionally, they eliminate the need for users to remember passwords or carry physical tokens, enhancing convenience and user experience. 

 \subsubsection{Location factors}

 The use of location as an additional authentication factor has been introduced recently \cite{article_13}. 
 This factor considers the individual's physical presence and compares it to their expected or usual location. If the user's location deviates significantly from their regular pattern or appears suspicious, it can trigger additional verification steps. 
 The prevalence of mobile devices and technological advancements like Global Positioning System (GPS) \cite{gps_location_factor}, IP geolocation \cite{ip_geolocation_factor}, and proximity authentication through ambient sounds \cite{ambient_location_factor} have led to the increased prominence of location-based authentication. 

\subsection{Emerging authentication factors}

Traditional authentication factors possess certain limitations, especially when employed for machine-to-machine (M2M) authentications. Hence, they pose a challenge in ensuring secure M2M communications and transactions. Machines, unlike humans, cannot produce and provide authentication factors like password or biometric data. New types of authentication factors, such as historical data, PUF and firmware integrity, are emerged 
to address these issues.  

\subsubsection{Historical data}

Historical data \cite{inproceedings_16} (e.g., previously exchanged sensor data between a client and server) is recently introduced as 
a strong authentication factor due to its dynamic nature 
and high leakage resilience. It is incorporated by selecting a set of indices from the historical dataset that is continuously expanding and using the corresponding data and tags to compute the response as part of the authentication process. As historical data is constantly expanding (even if an adversary compromises the data), it will get outdated 
soon after, leaving limited room for exploitation.  

\subsubsection{Physically unclonable functions (PUF)}

PUF \cite{article_14, inproceedings_80} utilise the physical distinction of each electronic device to generate a unique response when provided with a challenge. These distinctions are typically introduced during the manufacturing processes. Given the difficulty in replicating the same response, PUF is used as an authentication factor. While there may be slight variation in the response due to external factors, it can be rectified using fuzzy extractors.	 

\subsubsection{Firmware integrity}

Firmware integrity \cite{article_15} utilise the IoT device's firmware to verify its integrity and authenticity. Unlike traditional computing devices, most IoT devices have firmware as their operating system and interact directly with the hardware to complete the required assignments. With technological advancement, some IoT devices are installed with embedded Operating Systems (OS) such as Windows CE, embedded Linux OS, etc., where the image files are typically stored together with other essential files in a flash or embedded multi-media card. Any unauthorised modification would affect the integrity of the stored content, simplifying the authenticity checks of the IoT device by verifying the firmware integrity.

\section{Literature review on MFA protocols}\label{literature_review}

Before we dive into our security analysis of the vulnerable MFA protocols in Section \ref{security_analysis}, we would like to provide a summary of the literature reviews we conducted in several MFA protocols. 
Because, the literature review helps to better comprehend the current 
trends and advancements in MFA security. With the multitude of MFA protocols published, it is necessary to understand the approaches undertaken by researchers when designing the MFA protocols.  

In recent years, extensive research and development efforts have resulted in the development of numerous MFA protocols tailored to specific domains, including but not limited to cloud computing, finance, healthcare, and the rapidly expanding field of the internet of things (IoT). Numerous research papers have also been published to analyze and compare MFA protocols to identify and rectify security weaknesses and flaws, ultimately improving the overall security of MFA protocols. To identify vulnerabilities in MFA protocols, we reviewed 
several research papers in different domains. 
Due to space limitation, we discuss below only the selected ones that we believe are more relevant in each respective domain. 
Out of which, we identify vulnerabilities in ten of them, which are discussed 
in detail in Section \ref{security_analysis}.  

\subsection{Generic client-server architecture}\label{client_server_architecture}

Ertan et al. \cite{article_55} proposed a two-factor authentication protocol (using TOTP and PUF as the first and second authentication factors) to alleviate the security posture of TOTP systems. The MFA protocol 
uses PUF as storage to securely store the client’s secret. The authors assumed that Transport Layer Security (TLS) is in place to secure the communication channel between the entities.

Ivaylo Chenchev \cite{client_server_new1} proposed an MFA protocol secure communications in a client-server or peer-to-peer architecture. The protocol uses time-based onetime passwords (TOTP) and dynamically generated passwords (which is not stored anywhere) as its authentication factors. This protocol mainly intended to address the vulnerabilities of traditional password-based MFA protocols that potentially arise due to insecure generation of passwords.  

Chen et al. \cite{client_server_new2} proposed a biometrics-based three-factor authentication and key agreement scheme for multi-server environments. The authors mainly focused on addressing the weakness of conventional MFA protocols when adopted in a multi-server environment. This protocol employs user id, password and digital signature as authentication factors. Using this protocl, the authors claimed to achieve several impersonation and reply attacks.  

\subsection{Cloud computing}

As the adoption rate of cloud computing increases, adversaries targeting cloud services are also rising. MFA improves the robustness and efficiency of the authentication process, which directly enhances the overall security posture of cloud-based systems, safeguarding sensitive data from potential cyber threats. 

Bouchaala et al. \cite{article_18} proposed 
a cloud-based MFA protocol using user id and password, and additionally, smart card as authentication factors. The protocol is based on elliptic curve cryptography (ECC) and consists of four phases that include registration, authentication, key update, and card revocation. The key update phase facilitates the renewal of cryptographic keys stored in the smart card to maintain security, while the card revocation feature allows for the immediate revocation of compromised or lost smart card to reduce the likelihood of security compromise. Through the formal and informal security analysis conducted, the authors claimed that the MFA protocol is able to prevent attacks posed by adversaries.  

Lee et al. \cite{article_19} proposed 
a three-factor MFA protocol 
explicitly tailored for the cloud environment. The protocol employs password, smart device, and biometric as authentication factors. It consists of four essential phases, including the registration phase, login and authentication phase, password change phase, and identity update phase. Users are allowed to update their password and modify their personal information as needed. The authors proved that protocol was able to perform mutual authentication and establish a secured channel between both parties. 

Lee et al. \cite{article_20} proposed an innovative and improved three-factor authentication protocol to mitigate known attacks such as replay attacks, offline guessing attacks and Denial-of-Service (DoS) attacks. Besides password and biometric as the first and second authentication factors, the protocol allows the flexibility of using a laptop and/or smart card as the third authentication factor. Users can still complete the authentication process if the laptop or smart card is lost. The protocol consists of 
a registration and authentication phases. 

Otta et al. \cite{article_47} systematically surveyed MFA protocols explicitly tailored for securing cloud infrastructure. The paper covered the following key aspects: threats related to cloud authentication, analysis of the different types of authentication factors used in MFA, and a comparative analysis of MFA protocols designed by various researchers. The findings presented in the paper contributed to a structured approach in selecting the authentication factor, which can thwart impersonation attacks based on its uniqueness. 

Google developed Google Authenticator to provide its users with an additional layer of security \cite{Google_17}. With the enhancement, besides furnishing their username and password, users are required to furnish the OTP generated by the Google Authenticator registered to their account as the second authentication factor to verify their identity when accessing Google services. Any other services or applications that support TOTP algorithm can also utilise Google Authenticator as the second authentication factor. 

\subsection{Finance}

MFA 
is also widely adopted by financial sectors, such as banks, insurance companies, and various payment systems, 
to securely authenticate individuals before allowing them to perform any financial transactions. Scam and fraud transactions, and privacy of clients are some of the critical security concerns in this domain. 

Hassan and Shukur \cite{inproceedings_21} proposed a three-factor authentication protocol to enhance the authentication process for an electronic payment system. It combines three authentication factors: password, biometric, and OTP. The protocol consists of three phases: registration, authentication and transaction phases, to safeguard the payment system against a wide range of attacks. 

Durairaj and Ramachandran \cite{article_22} proposed an innovative approach to their ECC-based authentication protocol. In addition to low entropy password, International Mobile Subscriber Identity (IMSI) of the user’s device, and fingerprint, Durairaj and Ramachandran proposed to include the feature extracted from voice print, processed using Mel Frequency Cepstral Coefficient algorithm (MFCC) as an additional authentication factor. They proved that the MFA protocol is able to establish secure communication channels between two entities after successful mutual authentication. 

Similarly, Abiew et al. \cite{article_23} also adopted an innovative low-cost approach using another authentication factor, Keystroke Dynamics, to improve the authentication process. Through the experiment conducted, they proved that the protocol is able to mitigate the vulnerability of ATM PINS to dictionary attacks. 

Aburbeian et al. \cite{mfa_ml1} proposed a protocol that integrates MFA and machine learning to secure financial transactions. The protocol involves two stages security. In the first stage, the protocol employs fingerprint and OTP as authentication factors to authenticate users. In the second stage, the protocol involves a machine learning layer, which employs facial recognition as a decisive and third authentication factor. This is to further enhance the security and robustness of the authentication process. The authors claimed to have achieved several security features with high accuracy.   

\subsection{Healthcare}\label{healthcare}

In healthcare, researchers have focused on developing MFA protocols particularly designed to protect lost of patient data, privacy breaches and 
unauthorised access to health-related information.  

Sabeeh and Yassin \cite{article_24} proposed a two-factor authentication protocol to authenticate the administrator in the healthcare system using password and SMS token as authentication factors. The components include an administrator, user data entry, and a healthcare centre and are based on typical authentication phases such as initialization, registration, and authentication phase. Scyther tool was utilised to assess the security properties of the MFA protocol and proved to be capable of resisting known attacks. 

Shamshad et al. \cite{article_25} proposed an improved ECC-based two-factor authentication protocol that employs password and smart card as authentication factors to enhance the authentication process. Besides the typical registration and authentication phases, the protocol includes a password change phase to facilitate the change of password when required. While the performance of the protocol was observed to be sub-optimal, they proved that it is capable of resisting known attacks that affect other more efficient MFA protocols. 

 Ali et al. \cite{article_26} discovered that the protocol of Barman et al. \cite{article_61} was vulnerable to various attacks, such as session key leakage, server impersonation, and user impersonation attacks. Then, Ali et al.  \cite{article_26} proposed an improved three-factor MFA protocol to address those security flaws in \cite{article_61}. This protocol employs password, smart card, and biometrics as its authentication factors. The protocol also involves a user revocation feature, which allows to revoke users in case of loss of their smart card. The authors claimed to achieve several security properties, such as key resilience against key leakage attacks, impersonation attacks, and known attacks.

Dhillon et al. \cite{article_57} proposed a three-factor authentication protocol (involving password, biometrics, and smart card as its authentication factors) to establish a secure communication channel between a medical professional and a cloud server. The authors' informal and formal security analysis demonstrated that the proposed MFA protocol is capable of resisting various known attacks.

\subsection{IoT} 

The rapid proliferation of IoT devices has contributed to the surge in designed MFA protocols. The authentication process for IoT devices is unique due to resource-limited devices and M2M communication scenarios where human involvement is minimal or absent. That poses challenges for traditional authentication factors like PINs, passwords, and biometrics. Within the IoT domain, different subdomains such as Healthcare IoT, industrial IoT (IIoT), smart cities/homes, and wireless sensor networks (WSN) require specifically tailored MFA protocols to address their respective requirements and challenges. To provide a more comprehensive and thorough analysis, we further classify the MFA protocols of this domain into the following sub-categories. 

\subsubsection{Generic IoT}\label{giot}

Cvetković et al. \cite{inproceedings_49} presented on the application of MFA protocols in the context of IoT. 
With the complexity of IoT architecture taken into consideration, the paper briefly discussed the various IoT security techniques that stood out, the security objectives of IoT, selected MFA protocols relevant to IoT, and security considerations specific to MFA implementation in the IoT. The findings presented in the paper contribute to understanding MFA protocols suitable for IoT environments and provide insights into the challenges and considerations involved. With the limitations faced, they highlighted the need for efficient resource utilization, lightweight security authentication protocols, and cryptographic algorithms customised for the IoT environments. 

Halvor Vada presented a comparative analysis of various MFA protocols tailored for IoT systems. 
Vada highlighted the security challenges inherent in IoT systems at each layer of the selected 3-layer architecture and emphasized the importance of a robust authentication protocol to enhance security. The paper offered valuable insights into the implementation of MFA protocols for IoT systems by comparing a range of MFA protocols based on authentication factors employed, performance, strengths, and weaknesses. 

Chen et al. \cite{article_15} introduced a novel authentication protocol designed specifically to enhance the security of IoT devices through MFA. This protocol incorporates a hierarchical architecture based on the traditional IoT system, comprising the perception, network, and application layers. In addition to a secret cryptographic key and PUF, the integrity of the device’s firmware, which is used to ensure integrity, is also employed as one of the authentication factors. The protocol also supports firmware updates, enabling the deployment of patches and security enhancements. 

Mirsaraei et al. \cite{article_27} proposed an ECC-based MFA protocol on a blockchain platform for generic IoT devices. It employs password, biometric, and smart card as the authentication factors. In addition to the typical registration phase, login phase, and authentication phase, an update phase was included to facilitate password and biometric information updates. Through the informal and formal security analysis, the authors proved that the proposed MFA protocol satisfies the typical security requirements and is capable of resisting known attacks. Automated Validation of Internet Security Protocols and Applications (AVISPA) tool was utilised to automate the formal security analysis of this protocol. 

Zahednejad  et al. \cite{article_28} proposed a two-factor authentication protocol for generic IoT devices in an M2M context. The protocol employs a long-term cryptographic key and data items as authentication factors. The three phases include initialization, authentication, and a revocation phase to revoke compromised or lost devices. To simulate real-life scenarios, they assumed a strong adversary capable of compromising the server and retrieving all the information within and proved that the MFA protocol was able to ensure a secure authentication process and repel attacks. 

Sadhukhan et al. \cite{article_56} proposed an ECC-based lightweight three-factor authentication (involving user identity, password and bimetric data as the authentication factors) protocol for remote users in IoT network. The authors performed formal and informal security analysis to demonstrate that their proposed MFA protocol achieves several security features, such as resilience against impersonation and DoS attacks, perfect forward secrecy (PFS), and user anonymity, among others. 

Melki et al. \cite{article_58} proposed a lightweight ECC-based two-factor authentication protocol to provide secure communication channels between an IoT device and a gateway. 
A secret session identifier $ID_s$ (that is derived from a PUF output value) and a secret channel-based parameter $\sigma_i$ are used as authentication factors. The authors claimed to have achieved several security features, such as PFS, session secrecy, user privacy, and resilience against reply, side-channel and man-in-the-middle attacks. 

\subsubsection{Healthcare IoT}\label{hiot}

Jia et al. \cite{article_29} proposed a two-factor authentication protocol for a fog-driven IoT healthcare system using password and smart card as authentication factors. This protocol designed is specifically for healthcare applications in the IoT domain. It involves several phases, including system setup, registration of users and fog nodes, authentication, and key agreement. In addition, password updates, user revocation and re-registration, and fog node revocation were also included to ensure comprehensiveness. The bilinear pairing was utilized to compute cryptographic operations efficiently. Formal and informal security analyzes against known attacks were also performed to demonstrate the security of the MFA protocol. 

Al-Saggaf et al. \cite{article_30} introduced an innovative and improved two-factor authentication protocol specifically designed for the IoT-enabled healthcare ecosystem. The protocol combines smart card and biometrics as authentication factors. Given the challenges in post-quantum computing, the authors utilized a Post-Quantum Fuzzy Commitment scheme (PQFC) to protect the biometric template. It consists of several key phases, including setup, registration, login, authentication, and biometric revocation. By incorporating biometric revocation capabilities, the use of suspected compromised biometrics is prevented. 

Azrour et al. \cite{article_31} designed a two-factor authentication protocol to address the vulnerabilities observed in a prior work \cite{article_32}. This protocol employs password and smart card as authentication factors to secure communications between healthcare systems in the cloud IoT. It encompasses several phases, including system setup, registration of new sensors and users, login and authentication, and password update. Formal security analysis was performed using the Scyther tool and the ROM model, in addition to informal security analysis to verify the security of the MFA protocol. 

Chen et al. \cite{mfa_iomt} proposed a privacy-preserving three-factor MFA scheme to secure a cloud-assisted medical IoT. The protocol employs user id, password and biometric data as the authentication factors. The authors claimed to have achieved several security properties, such as post-quantum security, PFS, user anonymity, and resilience against reply and impersonation attacks. 

\subsubsection{IIoT}\label{iiot}

A wide range of industrial IIoT-based MFA protocols has been proposed over the years \cite{article_33, article_34, article_35, article_36, inbook_37, article_38, article_39}.  Sain et al. \cite{inproceedings_46} provided an overview of security issues related to Cyber Physical Systems (CPS) and explored the use of MFA to improve security. The authors also discussed the evolution of authentication, elaborated on the importance of MFA, and highlighted the rapid adoption of biometrics as one of the authentication factors. MFA protocols designed for CPS were deliberated and compared against an established set of evaluation criteria. 

Zhang et al. \cite{article_36} proposed a blockchain-based MFA protocol designed specifically for cross-domain IIoT systems. This protocol combines a long-term cryptographic key and PUF as authentication factors to enhance security. The protocol encompasses various phases, including registration, intra-domain authentication, cross-domain authentication to ensure a secure authentication process. A formal security analysis using BAN logic was performed to prove the security of the MFA protocol. 

Khalid et al. \cite{article_34} introduced an MFA protocol designed for cross-platform IIoT systems. This protocol combines password, smartcard, and biometric authentication factors to enhance the authentication process. The protocol utilizes the AES-ECC algorithm to secure communications between the entities. The protocol encompasses various phases, including setup, user and fog node registration, login, and authentication, ensuring only access to authenticated users. BAN logic was employed to perform a formal security analysis to prove the security of the MFA protocol.  

The three-factor authentication protocol proposed in \cite{article_35} employs a combination of authentication factors, including password, smart card, and biometric, to ensure secure communication channels. The protocol consists of seven phases: offline sensing device and user registrations, login, authenticated key agreement, biometric and password update, and dynamically sensing device addition and revocation. The security of the MFA protocol was proven using Real-Or-Random model (ROR).  

A newly designed low-interactivity Multi-Factor Authenticated Key Exchange (MFAKE) protocol named Secure Remote Multi-Factor protocol was introduced in \cite{article_36} and aims to enhance the security of M2M communication within the IIoT. The protocol ensures a robust authentication process by leveraging authentication factors such as password, biometric, and OTP. The key exchange was proven to be secured using the Bellare-Pointcheval-Rogaway model. 

Liu et al. \cite{inbook_37} introduced an innovative two-factor authentication protocol to enhance security in M2M communication within the IIoT landscape. The protocol employs long-term private keys and big data tags as authentication factors. The protocol begins with an initialization phase where the entities generate their respective private keys associated with the big data tags. The IoT device and the server need to provide evidence or proof that they possess knowledge of the data and its associated tag before establishing a secure connection between the two during the authentication phase. They demonstrated that the MFA protocol can achieve Key Compromise Impersonation (KCI) and Server Compromise Impersonation (SCI) Resilience, which are critical in IIOT.  

Jin et al. \cite{article_38} proposed a historical data-based multi-factor authenticated and confidential channel establishment (HMACCE) protocol for the M2M communication within the IIoT. This protocol involves a client and a server as the entities, with the client possessing a long-term symmetric key and a secret key as authentication factors. In contrast, the server holds a long-term symmetric key, historical data, and data tags. The protocol encompasses three phases, including initialization, tag generation, and online authentication and key exchange. During the tag generation phase, authentication tags are generated using the historical data and data tags, enabling subsequent verifications. They also proposed another HMACCE protocol named $\pi_{FS}$ to address the issue of adaptive bounded leakage.  

Cui et al. \cite{article_39} introduced a three-factor authentication protocol to satisfy the established security requirements of IIoT environments. The protocol incorporates three authentication factors, including password, biometric, and smart card. It encompasses various phases such as server initialization phase, smart device registration phase, user registration phase, user login phase, authentication and key agreement phase, password and biometric update phase, smart devices addition phase, and user revocation phase. All smart devices and users must be registered before communications are allowed. They demonstrated the security of the MFA protocol through informal and formal security analysis using ROR model. 

Han et al. \cite{mfa_iiot_new1} proposed a a three-factor MFA and key agreement protocol to secure the communication in IIoT that specifically involves three entities, namely user, gateway and sensing device. The protocol employs user id, password and biometric key (that is generated from the users biometric data using a fuzzy extractor) as its authentication factors. The protocol is mainly designed to address the vulnerabilities of a prior work \cite{mfa_iiot_new2}, such as 
lack of forward security and 
vulnerability to insider and session specific temporary information (KSSTI) attacks.   

\subsubsection{Smart cities/homes}

The two-factor authentication protocol for smart cities proposed in \cite{article_40} incorporates password and smart card as authentication factors. The protocol consists of several phases, including setup, user and sensors registration phase, login phase, authentication and key exchange phase, update phase, and revocation phase to allow for the removal of user privileges if necessary. Besides informal security analysis using BAN logic, which was typically used for formal security analysis, AVISPA tool was employed to prove the security of the MFA protocols. A comparison with other related MFA protocols was also conducted to demonstrate its effectiveness.  

Wang et al. \cite{article_41} designed an improved ECC-based three-factor authentication protocol based on ECC for smart homes using password, smart card, and biometrics as authentication factors. The initialization, registration, login and authentication, and password update phase were involved in the authentication process. They demonstrated the security of the MFA protocol through formal security analysis using BAN logic and demonstrated its efficiency through performance comparison with other related MFA protocols. 

The two-factor authentication protocol proposed in \cite{article_42} was designed to address the security issues observed in \cite{article_43}. Password and mobile device are employed as authentication factors in the MFA protocol. The protocol involves a series of phases, from initialization and device registration to authentication and key agreement processes. The protocol also supports password updates for enhanced security. Secure communication and interaction between the Mobile User and the Smart Device are established through the Home Gateway. A combination of tools, including BAN logic, ROR model, and AVISPA tool, was utilised to demonstrate the security of the MFA protocol. 

\subsubsection{WSN}\label{wsn}

Wang et al. \cite{article_50} aims to investigate and understand the failures observed in the security proofs of MFA protocols specifically designed for mobile devices. The paper provided an overview of MFA protocols designed for mobile devices, an understanding of security proofs failures in MFA protocols, the development of an enhanced set of evaluation criteria, and an analysis of a selection of ten MFA protocols. They also proved that protocols with formal security proofs were able to better satisfy the established evaluation criteria which aid in the design of a more secure MFA protocol for mobile devices. 

The three-factor authentication protocol designed by authors in \cite{article_44} for WSNs consists of three phases, including registration of users and sensors phase, login and authentication phase, and password change phase. It employs password, biometric, and smart card as authentication factors. All users and sensors must register with the gateway before establishing any connection. The protocol utilizes "honey list" and "fuzzy extractor" techniques to enhance security . ProVerif tool was employed to prove the security of the MFA protocol. The protocol's performance also proved to be exceptionally better compared to other related MFA protocols. 

Kumar et al. \cite{article_45} proposed a three-factor authentication protocol incorporating password, smart card, and biometric as authentication factors. They incorporated several phases, including the initialization base station phase, user and sensor node registration phase, login and authentication phase, and password and session key updating phase, into the protocol design where users can update their password and generate new session keys for enhanced security. The protocol also allowed the inclusion of new sensor nodes into the network with proper authentication procedures through the adding new nodes phase. The ROR model and the AVISPA tool were employed to demonstrate the security of the MFA protocol, which proved to satisfy the security requirements. 

\subsection{Mobile authenticators}

Recently, mobile applications are widely used as authentication factors in most of the domains mentioned above. For example, Google developed Google Authenticator to provide its users with an additional layer of security \cite{Google_17}. With the enhancement, besides furnishing their username and password, users are required to furnish the OTP generated by the Google Authenticator registered to their account as the second authentication factor to verify their identity when accessing Google services. Any other services or applications that support the TOTP algorithm can also use Google Authenticator as the second authentication factor. 

\begin{figure}[htp]
    \centering
    \includegraphics[scale=0.5]{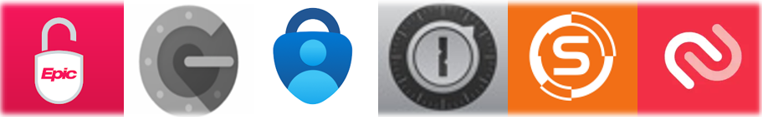}
    \caption{Affected authenticator applications}
    \label{fig:image2}
\end{figure}

However, such authenticators are also not without security risks. In a recent publication \cite{inproceedings_11}, researchers discovered vulnerabilities in several mobile authenticators that exposed the unique secret key. The affected authenticators (as illustrated in Figure \ref{fig:image2}) includes Epic Authenticator, Google Authenticator, Microsoft Authenticator, Sophos Authenticator, Red Hat Free OTP, and Twilio Authy Authenticator. These vulnerabilities allowed an adversary to access the unique secret key stored plainly at specific repository locations, such as directories or database files. In addition, the unique secret key can also be retrieved from memory during specific periods. 


\section{Adversarial model}\label{threat_model}

\begin{figure}[htb!]
    \centering
    \includegraphics[scale=0.43]{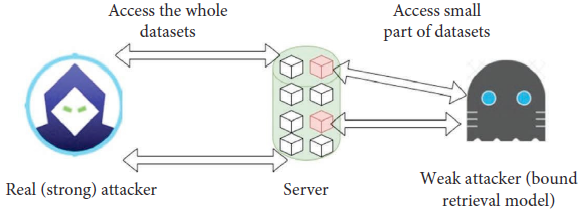}
    \caption{Snapshot of the comparison between a strong and weak attacker \cite{article_28}}
    \label{fig:image6}
\end{figure}

In order to assess the security of MFA protocols, it is essential to assume a realistic and concrete adversary model. In fact, one of the limitations of most MFA protocols is failing to assume strong and realistic adversaries. Therefore, it is crucial to consider a realistic adversary model. In our model, we consider the following capabilities of the adversary in an MFA context:
\begin{itemize}
    \item The adversary has full control of messages transmitted over the public channel, i.e., it can intercept, eavesdrop, and redirect it. 
    \item The adversary can acquire design of the proposed MFA protocol.
    \item The adversary can acquire the first authentication factor, e.g., password.
    \item A strong adversary can obtain all the data from the device if he gets access to the device (cf. Figure \ref{fig:image6} for the comparison between weak and strong adversaries).
    \item In case of PFS attack, we assume that the adversary can obtain the long-term secrets of both parties.
\end{itemize}

\section{Security analysis of MFA protocols }\label{security_analysis}

As highlighted in the introduction, we perform 
critical security analysis on several MFA protocols to identify potential security flaws. Of these, we identify 
serious security flaws in ten of the protocols. In this section, we provide a detailed discussion of the 
vulnerabilities identified in these protocols. In addition, since the runtime performance of the protocols is critically important, especially in resource-constrained devices and/or hard real-time constrained systems, we also present a comparative performance analysis of the protocols to assess their efficiency. 

\subsection{Evaluation criteria}\label{evaluation_criteria}

To assess the security of an MFA protocol, it is essential to establish a set of evaluation criteria. 
Different researchers adopted different sets of security evaluation criteria considering various domains and contexts. After analyzing the criteria in the existing frameworks, we introduce some new criteria, such as distinctiveness of authentication factors, independence of authentication factors, and leakage resilience, to form our set of evaluation criteria. In sum, it comprise 8 criteria, which are briefly discussed and summarized in Table \ref{table_evaluation_criteria}. We believe these criteria are 
very essential 
to evaluate MFA protocols in different domains, including the multi-factor authenticated key exchange (MAKE) domain in IoT settings. 

\begin{table*}[htb!]
\centering
\caption{Evaluation criteria}
\label{table_evaluation_criteria}
\begin{tabularx}{\linewidth}{llX}
\specialrule{.13em}{.05em}{.05em}
S/No. & Evaluation criteria &
  Description \\
  \hline
C1 & Mutual authentication & 
Both parties must verify each other's identities before 
advancing the authentication process and establishing a session key. \\
C2 & Distinctiveness of factors & Employ distinct authentication factors each selected from different categories, such as knowledge factors, possession factors, inherent factors, historical data, etc. 
\\ 
C3 & Independence of factors & Ensure each authentication factor is independent from other. For example, the generation of one factor must not depend on any other factor. Similarly, one factor must not be protected (e.g., encrypted) by using any other factor.\\
C4 & Leakage resilience & 
Ensure that any data 
leakage cannot compromise keys 
or authentication factors. In addition, keys and authentication factors must be computationally infeasible to be predicted or guessed. \\
C5 & Perfect-forward secrecy (PFS) & 
Ensure that the leakage of long-term keys (client or server) cannot compromise the security of previous sessions.\\
C6 & User anonymity & 
Preserve user's identity 
during the authentication process. \\
C7 & Resilience against known attacks & Protect the authentication process 
against known attacks, such as MITM, replay, password-guessing, impersonation, insider, and DoS attacks. \\
C8 & Adversary assumption & Assume strong and realistic adversaries who possess adequate skills and resources to perform sophisticated attacks. \\
  \toprule
\end{tabularx}
\end{table*}

\subsection{Security analysis}

As discussed in the preceding sections, we identified vulnerabilities in ten of the MFA protocols we analyzed. In this section, we discuss a detailed account of the vulnerabilities identified. 

\subsubsection{Vulnerable MFA Protocol 1 \cite{article_38}}\label{vuln_proto_1}

\begin{figure*}[htp]
    \centering
    \includegraphics[scale=0.9]{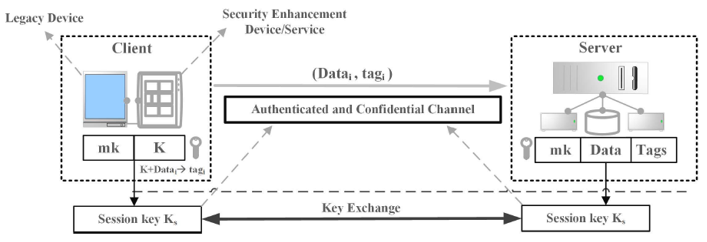}
    \caption{Overview of the HMACCE Protocol \cite{article_38}}
    \label{fig:image4}
\end{figure*}

As discussed in Section \ref{iiot}, Jin et al. \cite{article_38} proposed a historical data-based multi-factor authenticated and confidential channel establishment (HMACCE) protocol for the M2M communication within the IIoT, as illustrated in Figure \ref{fig:image4}. This protocol involves a client and a server as the entities. The client possesses a long-term symmetric key and a tag generation secret key as first and second authentication factors while the 
server possesses a long-term symmetric key, historical data and data tags as first, second and third authentication factors, respectively. 
The protocol encompasses three phases, including initialization, tag generation, and online authentication and key exchange. During the tag generation phase, authentication tags are generated using the historical data and data tags, enabling subsequent verifications. 
The authors claimed to address the key leakage issues of a prior historical data-based two-factor authentication protocol \cite{inproceedings_59}. This protocol comprises two versions, namely $\pi_{woFS}$ and $\pi_{FS}$, that are designed without and with forward secrecy, respectively. The latter is also claimed to address the issue of adaptive bounded leakage. 
 
Based on our analysis, this protocol exhibits several vulnerabilities: 1) The authors claimed that adding an additional authentication message will ensure mutual authentication between the client and server. Based on the authentication steps provided, it is true that the client can verify the authentication message, $M = h(mk || Y || sid || Auth’)$, sent by the server. 
However, further checks on the MFA protocol revealed that the client does not send an explicit authentication message to the server. Hence, the server cannot verify authenticity of the client, therefore failing the mutual authentication criteria \emph{C1}. 2) Both of the client's authentication factors, i.e., the symmetric authentication key $mk$ and the 
tag generation secret key $K$, are in the same authentication factor category, 
i.e., possession factors. Given that it does not fully satisfy the intent of MFA, where multiple authentication factors from different categories should be used to prove identities \cite{misc_60}, 
it fails the authentication factors distinctiveness criteria \emph{C2}. 3) The second authentication factor $t_i$ of 
this protocol is protected using the first authentication factor $sk^{1}_{ids,idc}$ during transmission, hence failing 
the authentication factors independence criteria \emph{C3}. 
4) During the tag generation phase, in which the client transfers a piece of data $d_i$ to the server, the server computes an authentication tag $t_i$ using $K$, which is the second authentication factor of the client. The equation used to derive $t_i$ is: $$t_i= K .h(d_i  || i)+k_i  (mod p)$$ As a result, the computation of the tag $t_i$ depends on the value of the second authentication factors of the client and the server. Any compromise to the second authentication factor could potentially lead to the compromise of the third authentication factor, hence failing the criteria \emph{C3} again. 5) This protocol is also susceptible to data- and tag-stealing attacks in which adversaries could retrieve all the historical data and tags. Since data and tags are used as authentication factors, such leakage can also compromise the authentication factors and the session keys. Therefore, it fails the leakage resilience criteria \emph{C4}. 6) The entropy of its sensor data (which is used as authentication factor) is between 4.52 and 7.80, which is low and could easily be predicted. Hence, it fails the criteria \emph{C4} again. 7) This protocol does not employ any client anonymity mechanism, thus failing criteria \emph{ C6}. 8) The authors assume a weak adversary (called a bounded-retrieval model) who can access only a fraction of the historical data after he compromises the server. This is an unrealistic assumption and fails the criteria \emph{C8}. On the other hand, the second version (i.e., $\pi_{FS}$) of this protocol satisfies PFS while the first version (i.e., $\pi_{woFS}$) does not. 

\subsubsection{Vulnerable MFA Protocol 2 \cite{article_27}}

As discussed in Section \ref{giot}, Mirsaraei et al. \cite{article_27} proposed an ECC-based MFA protocol on a blockchain platform for generic IoT devices. It employs password, biometric, and smart card as the authentication factors. 
The authors conducted an informal and formal (e.g., using the AVISPA tool) analysis and claimed that their protocol is capable of satisfying a set of established security requirements and resisting attacks including key leakage, MITM attacks, DoS attacks, etc. 

We perform further analysis in this protocol to identify potential security flaws. In their security analysis, it was assumed in ``Assumption 5'' that \textit{“the malicious attacker can obtain only one parameter in an equation at a time.”}. It is crucial to note that adversaries can employ various sophisticated techniques and strategies to acquire multiple parameters simultaneously. Given that a strong adversary could potentially acquire the information 
discussed in Section \ref{threat_model}, there is a risk in this protocol that the session key can be computed and compromised. The session key is computed using the following equation: $$SK = M_2 \oplus V_1$$
Using the information acquired 
discussed in Section \ref{threat_model}, the adversary can easily obtain the values of $M_2$ and $V_1$ to compute $SK$. This is because, $M_2$, which is transmitted through the insecure channel from the server to the client, can be easily retrieved by eavesdropping. $V_1$ can also be computed using the following equation: $$GID_{i} = ID_{i} \oplus V_{1}$$

Similarly, $GID_i$ is transmitted through the insecure channel from the client to the server and can be retrieved by the adversary through eavesdropping. As for $ID_i$, the adversary could acquire the information through other forms of attacks such as phishing, keylogging, malware, etc. With the computed session key, the adversary would be able to decipher the information transmitted via the secured communication channel using the compromised session key. Therefore, this protocol fails to achieve several evaluation criteria, including C4, C5, C7 and C8. 


\subsubsection{Vulnerable MFA Protocol 3 \cite{article_28}}

As highlighted in Section \ref{giot}, Zahednejad  et al. \cite{article_28} proposed a two-factor authentication protocol for generic IoT devices in an M2M context. The protocol employs a long-term cryptographic key and historical data items as authentication factors. 
To simulate real-life scenarios, they assumed a strong adversary capable of compromising the server and retrieving all the information within (as illustrated in Figure \ref{fig:image6}) and proved that the MFA protocol was able to ensure a secure authentication process and repel attacks. They also employed ROR model to to formally prove how their protocol achieves perfect forward secrecy and key compromise resilience. 


We perform a further analysis of the protocol to assess its security. As per the threat model, the adversary can only obtain $N-1$ authentication factors and is able to eavesdrop any information passed through the public communication channel. Based on that assumption, it is possible to acquire the values of $R1$, $R2$, $Y$ and $TID_c$, which are transmitted in plain through the public channel. The adversary would also be able to acquire the value of one of the authentication factors $mk$. Given that $spk_s$ is the server’s public key, it is publicly available and can be easily acquired by the adversary. With the acquired information, the adversary would be able to compute $r1$, $r2$, and $X$ using the following equations:
\begin{enumerate}
    \item $R1 = mk \oplus r1$;
    \item $R1 = mk \oplus r1$;
    \item $X = Y - H(r1||r2)$;
\end{enumerate}

Using the acquired and computed values, the adversary can compute the value of the session key $SK_s$ and decipher the information that is transmitted via the communication channel secured by the session key $SK_s$ using the equation:$$SK_s = H(mk|r1|r2|X|TID_c|spk_s)$$. Moreover, the construction of the authentication process in this protocol is entirely dependent on the first authentication factor (i.e., mk). Meaning, if the adversary manages to get mk, he can simply forge the whole authentication process. Therefore, this protocol does not even satisfy multi-factor authentication, and it fails several evaluation criteria, such as C3, C4, C5, C7 and C8. 

\subsubsection{Vulnerable MFA Protocol 4 \cite{article_45}}

As discussed in Section \ref{wsn}, Kumar et al. \cite{article_45} proposed a three-factor authentication protocol for wireless sensor networks. 
It employs password, smart card, and biometric as authentication factors. 

However, based on our analysis, this protocol also has several limitations. First of all, 
a strong adversary was not considered in their threat model. In the scenario whereby a strong adversary is able to intercept the information that passes through the open communication channel, i.e. $ID_{sn}^i$, $N_{ur}$, $h(RN_{sc})$, $TS_1$ and $TS_5$, and obtain $N-1$ authentication factors, i.e., user id and password, and smart card, the session key used by the entities can be computed. Using the data acquired, the adversary can compute the value of the session using the equation:
$$K_{ss} = h(ID_{ur}||ID_{sn}^i||U_{rg}^i||h(ID_{ur}||N_{ur})||h(RN_{sc})||TS_1||TS_5)$$

It is possible to infer that the value $TS_5$ is meant for a particular session key by checking the value of $l_{10}$, which is equal to $(h(K_{sh} \oplus U_{rg}^i)||ID_{ur})$. Based on Kerckhoffs’s principle \cite{kerckhoffs_principle} and that hash values have a fixed length, the adversary is able to derive the length of $h(K_{sh} \oplus U_{rg}^i)$. Using this information, the adversary can extract the value of $ID_{ur}$ by removing the bits belonging to the hash value and determine if $TS_5$ is meant for the affected user that is attempting to establish a secure communication channel, which results in security and anonymity breaches. 
Overall, this protocol fails several evaluation criteria, such as C4, C5, C6, C7 and C8. 

\subsubsection{Vulnerable MFA Protocol 5 \cite{article_31}}

As discussed in Section \ref{hiot}, Azrour et al. \cite{article_31} designed a two-factor authentication protocol for healthcare IoT systems. The protocol was designed to address vulnerabilities 
of a prior work \cite{article_32}. This protocol employs password and smart card as authentication factors to secure communications between healthcare systems in the cloud IoT. 
In addition to the informal security analysis, the authors claimed that they performed a formal security analysis 
using the Scyther tool and the ROM model to verify the security of their MFA protocol. 

However, upon performing a security analysis, we discovered several flaws in this protocol. 
First of all, the authors did not provide a clear threat model 
nor indicated capabilities of the assumed adversaries. 
In addition, based on their formal and informal analysis, the authors claimed that they achieved perfect forward secrecy 
assuming that the values of $x_s$ and $MID$ are kept secret. However, our analysis of this protocol proved the other way round. 
$MID$ is transmitted plainly through the public channel; therefore, it is not secret. In addition, a strong adversary would be capable of acquiring the value of $x_s$ from the cloud server through alternative means. Then, the value of $w_i$, which is used to compute the session key $S_{key}$, can be obtained using the following equation: $$w_i= h(MID||x_s)$$ 

$Id_{SN}$ is also transmitted plainly through the public channel, therefore it can be eavesdropped. Using the values $w_i$, $MID$ and $Id_{SN}$, the adversary can compute the session key $S_{key}$ using the equation: $$S_{key}= h(w_i||MID||Id_{SN})$$

Therefore, this protocol fails several evaluation criteria, including C4, C5, C6, C7, and C8. 

\subsubsection{Vulnerable MFA Protocol 6 \cite{article_26}}

As discussed in Section \ref{healthcare}, Ali et al. \cite{article_26} proposed a three-factor symmetric key-based secure AKA protocol for Telecare Medicine Information Systems (TMIS). It was designed to address vulnerabilities of a prior work \cite{article_61}. The protocol involves password, smart card and biometrics as its authentication factors. A revocation/re-register phase was also incorporated to enable the revocation of users 
in case of loss of a smart card. Through the informal and formal analysis, the authors demonstrated that the MFA protocol is 
resilient against key leakage, impersonation and known attacks, unlike several other authentication protocols. 

However, our analysis proves that the proposed protocol fails to achieve several security properties, as discussed follows. The proposed MFA protocol computes its session key $SK_{ij}$  as follows: $$SK_{ij} = h(Y_{RC}||SID_j||T_3) == SK_{ij}^{'} = h(Y_i||SID_j||T_3)$$ 
Through our analysis, 
the values of $SID_j$ and $T_3$  can be intercepted from the open communication channel. It is also possible to compute the value $Y_i$ using the following equation.  $Y_i = h(SID_j||HID_i^{'}||R_{rand2}||T_1)$, where:
\begin{itemize}
    \item $HID_i^{'}= h(PID_i)$ (acquire $PID_i$ by compromising the first authentication factor);
    \item $R_{rand2} = R_{rand2}^{'} \oplus ID_i$ (acquire $R_{rand2}^{'}$  by intercepting the message from the open communication channel);
    \item $T_1 = T_1^{'} \oplus HID_i^{'}$ (acquire $T_1^{'}$ by intercepting the message from the open communication channel);
\end{itemize}
Upon acquiring the required information 
as shown above, the adversary can compute the value of the session key $SK_{ij}$ and decipher the information that is secured using the session key. 
The authors also assumed that the adversary cannot steal the private key of the registration center, which is a weak assumption as such keys can be leaked using side-channel or other techniques. 
Therefore, this protocol 
does not also achieve key leakage resilience (C4), perfect forward secrecy (C5), 
resilience against known attacks (C7), and strong adversarial assumption (C8). 


\subsubsection{Vulnerable MFA Protocol 7 \cite{article_55}}

As discussed in Section \ref{client_server_architecture}, Ertan et al. \cite{article_55} proposed a two-factor authentication protocol for a generic client-server architecture. It aimed to alleviate the security posture of TOTP systems. It employs TOTP and PUF as its first and second authentication factors. The MFA protocol 
uses PUF as storage to securely store the client’s secret. The authors assumed that 
the communication channel between the entities is secured using TLS. 

Based on our analysis, it is observed that the paper did not explicitly demonstrate the complete authentication process, specifically mutual authentication, which is an important requirement in MFA protocols. The client sends the following messages to the server for the authentication:

\quad a) $M_1= H(k_1, (t-t_0)/I) \oplus r$;

\quad b) $M_2= H(k_2 \oplus r, (t- t_0)/I)$;

\quad c) $c_1$, and;

\quad d) $c_2$.

The server computes the value of $H(k_2 \oplus r, (t-t_0)/I)$ and verify if the computed value is the same as $M_2$ to authenticate the client. There is, however, no clear indication of verification performed by the client to determine the identity and authenticity of the server. Hence, it does not provide mutual authentication. 
Furthermore, the authors did not consider other critical security requirements in MFA, such as key leakage resilience, PFS, user anonymity, resilience against known attacks, and strong adversarial assumption. Therefore, it fails the evaluation criteria C1, C4, C5, C6, C7, and C8. 

\subsubsection{Vulnerable MFA Protocol 8 \cite{article_56}}

As discussed in Section \ref{giot}, Sadhukhan et al. \cite{article_56} proposed an ECC-based lightweight three-factor authentication protocol for remote users in a generic IoT network. It employs user identity, password and bimetric data as its authentication factors.
Through their informal security analysis, the authors claimed to have achieved several security goals, including resilience against impersonation and DoS attacks, 
PFS, and user anonymity, among others. 

However, our further analysis on this protocol reveals that most of the claimed security goals are not properly achieved. For example, the authors claim of achieving PFS is just by assuming that the secret random number $R_{U}$ used in the session key generation cannot be compromised by an attacker even if the pre-shared encryption/decryption symmetric key $K_{X}$ is compromised by the attacker. This is a wrong assumption. If the attacker gets $K_{X}$ (which is the underlying assumption in PFS), he can intercept the packet sent over the public channel and decrypt it to obtain $R_{U}$ (see Fig.6 in \cite{article_56}). Because, $R_{U}$ is protected by only $K_{X}$. That is also how the gateway obtains $R_{U}$ in their protocol. Therefore, this protocol does not achieve PFS. 

In addition, the authors do not consider proper authentication factors as the user ID and biometric can be easily eavesdropped over the public communication channel. Moreover, the mutual authentication is based on the hash of the three factors, i.e., $H_{U} = h(ID_{U}||PW_{U}||B_{U})$. This value (alongside  $R_{U}$) is also encrypted using $K{X}$, i.e., $E_{Kx}(H_{U}||R_{U})$, and send to other parties over the public channel (see Figure \ref{fig:image9} or Fig.6 in \cite{article_56}). That means, it is completely dependent on the pre-shared encryption/decryption key $K_{X}$ (see the discussion about $R_{U}$ above). In other words, the security of the protocol is entirely dependent on  $K_{X}$, hence the authentication factors become meaningless. Therefore, this protocol does not achieve proper mutual authentication,  distinctiveness of factors, and independence of factors. 

Furthermore, the authors claim of user anonymity is by assuming that 
the user’s identity, i.e.,  $ID_u$, is never communicated in plain. However, the user sends a message, consisting of the variables $ID_u, E_{kx}(H_u||R_u), T_1$, to the IoT node over the public channel (as highlighted in Figure \ref{fig:image9}). 
Since $ID_u$ is sent in plain over the open communication channel, an adversary can obtain it by eavesdropping the open communication channel. Hence, the protocol does not preserve the user’s anonymity. 

In addition, this protocol does not properly achieve resilience against key leakage and known attacks nor provided strong threat model. Overall, this protocol fails all of our security evaluation criteria, i.e., C1 through C8.

\subsubsection{Vulnerable MFA Protocol 9 \cite{article_57}}

As highlighted in Section \ref{healthcare}, Dhillon et al. \cite{article_57} proposed a three-factor 
MFA and key exchange protocol for a healthcare system. In particular, it is designed to establish a secure communication channel between a medical professional and a remote patient monitoring in Cloud-IoT environments. The protocol involves a password, biometrics, and smart card as its authentication factors. The authors 
claimed to have achieved 
several security goals, such as mutual authentication, user anonymity, forward secrecy and many other known attacks. 

However, our further analysis on the proposed protocol revealed several security flaws. For example, even though the protocol is supposed to use three authentication factors, credentials computed based on the user's password (1st factor) and biometric feature (2nd factor) are made to be stored in the smart card (3rd factor). This means, the two factors are dependent on the safety and security of the 3rd factor, failing our factors independence criteria C3. 

In addition, the authors claim of achieving forward secrecy is infeasible. To achieve this, they are relying on the MP's identity ($ID_{MP}$), its private key $X$, and other parameters $c$ and $u$. First of all, $u$ is transmitted over a public channel (as seen in Fig.5 of \cite{article_57}), which can be easily intercepted. Secondly, the assumption in forward secrecy is that the adversary can compromise long-term credentials of the client and server, hence the authors cannot rely on $X$. Third, $c$ is computed based on the private key of the cloud server, which is the same issue as above. Fourth, the $ID_{MP}$ is also stored both in the client and cloud server, which can be obtained in the same way or in several other means. In fact, $ID_{MP}$ should not also be used for such purpose since it is not a secret factor. Therefore, this protocol fails the forward secrecy criteria C5. 

Furthermore, there is no resilience against session key or long-term key leakage attacks or other known attacks. The authors also did not put a strong adversarial assumption that compromise identities and credentials of the user and cloud server. Overall, this protocol fails several security criteria, including C3, C4, C5, C7 and C8.

\subsubsection{Vulnerable MFA Protocol 10 \cite{article_58}}

As discussed in Section \ref{giot}, Melki et al. \cite{article_58} proposed a lightweight ECC-based two-factor authentication protocol to provide secure communication channels between an IoT device and a gateway (see Figure \ref{fig_protocol10_auth}). The protocl uses a secret session identifier $ID_s$ (that is derived from a PUF output value) and a secret channel-based parameter $\sigma_i$ are used as authentication factors. The authors claimed to have achieved several security properties, such as PFS, session secrecy, user privacy, and resilience against reply, side-channel and man-in-the-middle attacks. 

However, our analysis of the protocol reveals several flaws of the protocol. 
First of all, the assumed authentication factors, i.e., $ID_s$ and $\sigma_i$, are not appropriate factors since such features are not reliable and prone to several security problems. These factors are not also distinct as both belong to the family of possession factors. 
In addition, the authors' assumption that "the adversary's probability of obtaining $ID_s$ is exceptionally low" is unrealistic and it diminishes the principal threat assumption of MFA, in which the attacker can achieve $N-1$ authentication factors. In fact, there are several ways to obtain $ID_s$, e.g., via side-channel techniques as it is stored in the client and server devices. 

Furthermore, the proposed protocol is susceptible to both client and server impersonation attacks. Hence, an adversary can potentially obtain the list of $ID_s$ from the gateway. In addition, the adversary can plant itself in between the IoT device and the gateway and establish separate communication channels with either of them. In that case, the IoT device would send 
$Message_{1}: <M_1,M_2, TS_A>$ to the adversary instead of the gateway. The adversary will attempt to retrieve the values of $R_A$ and $\tau_{i}$ by computing $M_{3}^{a} = h(ID_s||TS_A)$, $R_A = M_{3}^{a} \oplus M_{1}$, $\tau_{i} = M_{2} \oplus R_{A}$. The adversary will also compute $\sigma_{i}^{'} = Rep(N_{0,A^{a}}, \tau_{i})$ where $N_{0,A^{a}}$ is the channel-based nonce between the IoT device and adversary. The adversary then computes $Message_{1}^a$ containing $<M_{1}, M_{2}^{a}, TS_{A}>$ where $M_{2}^{a} = R_{A} \oplus \tau_{i}^{a}$. Here, $\tau_{i}^{a}$ is computed using the equation $Gen(N_{0,B^{a}}) = (\sigma_{i}^{a}, \tau_{i}^{a})$ where $N_{0,B^{a}}$ is the channel-based nonce between the adversary and gateway. Upon receiving $Message_{1}^{a}$, the gateway can compute the following values and send them back to the adversary as $Messsage_{2}: <M_4, M_5, TS_B>$ where:
\begin{enumerate}[label=\alph*),ref=\alph*]
    \item $SK=h(ID_s||TS_A||TS_B||R_A||R_B||\sigma_i^{a^{'}})$, where $\sigma_i^{a^{'}}=Rep(N_{0,B^{a}},\tau_{i}^{a})$; 
    \item $M_4=h(ID_s||TS_B||TS_A||R_A) \oplus R_B$, and;
    \item $M_5=h(SK \oplus R_A \oplus R_B)$.
\end{enumerate}
The adversary will then send $Message_{2}^{a}$: $<M_4, M_{5}^{a}, TS_B>$ to the IoT device where $M_{5}^{a}=h(SK_a \oplus R_A \oplus R_B)$ and $SK_a=h(ID_s||TS_A||TS_B||R_A||R_B||\sigma_i^{'})$. Upon receiving $Message_{2}^{a}$, the IoT device will compute the following values and sent back to the adversary as $Messsage_{3}: <M_8,TS_{A^{'}}>$ where  $M_8=h(SK_{a^{'}}||ID_s)$. Lastly, the adversary will send $Message_{3}^{a}: <M_{8}^{a},TS_{A'}>$ where $M_{8}^{a}=h(SK^{'}||ID_s)$ to the gateway. With the above actions performed, the adversary can successfully plant itself in between the IoT device and gateway to intercept all the transmitted information.

Furthermore, the authors claimed to have achieved PFS just by considering the one-way property of a hash function and the secrecy of the session identifier $ID_s$. This is also an unrealistic claim. Because, the $ID_s$ is not properly secure (as discussed above) and the other parameters (e.g., timestamp) are transmitted over the public channel (see Figure \ref{fig_protocol10_auth}). More importantly, the authors' assumption of PFS is different from the actual definition. In  reality, 
PFS is a critical security requirement in key exchange which requires that past session keys remain secure even if the long-term credentials of both client and server are compromised. Therefore, this protocol does not achieve PFS in many ways.    

In addition, the authors excluded a malware embedded in one of the communicating devices and an adversary present in the same subnet (as it can obtain channel-based parameters) from their threat assumption. 
These are unrealistic assumptions, which makes their adversarial model weak. Overall, this protocol fails to achieve any of our evaluation criteria, i.e., C1 through C8.

\subsection{Summary of the security analysis}

The analysis performed demonstrated that not all MFA protocols are without weaknesses or flaws. Table \ref{table_4} summarizes 
results of our evaluation on the ten protocols using our evaluation criteria. 
One of the key insights drawn from the analysis is that having more authentication factors does not necessarily translate to better security. As shown in the above analysis, most of the MFA protocols do not consider the critical security requirements discussed in our evaluation criteria.

Some MFA protocols were not designed with mutual authentication in mind \cite{article_66}. With the increase in scams \cite{misc_62}, it is no longer surprising that individuals require assurance that the entities they are interacting with are trusted and legitimate. Organizations verifying the authenticity of individuals alone are no longer sufficient, as they would need to prove their identity to the individuals interacting with them too.  

As shown in the above analysis, employing interdependent multiple authentication factors 
will negatively affect the security of the MFA protocol it is originally supposed to provide. Compromise in one of the authentication factors may result in a cascading effect leading to the compromise of the other authentication factors.

Privacy and anonymity are two features that people increasingly demand as they become more aware of the importance of safeguarding their personal information and identity. Moreover, there are privacy regulations \cite{misc_63, misc_64, misc_65} that organizations must comply with. Secure generation of session keys that are only accessible by authorized parties, therefore, becomes one of the key requirements \cite{article_45}, which seems to be lacking in some of the investigated MFA protocols. A combined attack such as MTIM attack to obtain information transmitted through the open communication channels and other attacks e.g., smart card loss attack, side-channel attack, or malware infection resulting in the loss of one authentication factors, a strong adversary is able to repeatedly compute the session key generated for each session and use it to decipher the sensitive information protected by the session key. 


Lastly, it is imperative to accurately depict real-life threat scenarios. The capabilities of adversaries and all possible entry points that could be targeted by them must be considered in the design of a robust, efficient, and secure MFA protocol. The weaknesses and flaws identified through this analysis is a clear indication of fundamental but critical security requirements that should have been considered when designing MFA protocols.

\begin{table*}[htb!]
\centering
\caption{Security evaluation results of the 10 MFA protocols}
\label{table_4}
\resizebox{\textwidth}{!}{%
\begin{tabular}{lllllllllll}
\toprule
Protocol & Domain & Authentication Factors & C1 & C2 & C3 & C4 & C5 & C6 & C7 & C8 \\
\midrule
Protocol 1 \cite{article_38} ($\pi_{woFS}$) & IIoT & LSK + TGK$^{c}$ + HD$^{s}$ + HDT$^{s}$ & $\times$ & $\times$ & $\times$ & $\times$ & $\times$ & $\times$ & $\checkmark$ & WA \\
Protocol 1 \cite{article_38} ($\pi_{FS}$) & IIoT & LSK + TGK$^{c}$ + HD$^{s}$ + HDT$^{s}$ & $\times$ & $\times$ & $\times$ & $\times$ & $\checkmark$ & $\times$ & $\checkmark$ & WA \\
Protocol 2 \cite{article_27} & Generic IoT & PW + SC + BD & $\checkmark$ & $\checkmark$ & $\checkmark$ & $\times$ &$\times$& $\checkmark$ & $\times$ & WA \\
Protocol 3 \cite{article_28} & Generic IoT & LSK + HD & $\checkmark$ & $\checkmark$ & $\times$ & $\times$ &$\times$& $\checkmark$ & $\times$ & SA \\
Protocol 4 \cite{article_45} & WSN & PW + SC + BD & $\checkmark$ & $\checkmark$ & $\checkmark$ & $\times$ &$\times$& $\times$ & $\times$ & WA \\
Protocol 5 \cite{article_31} & Healthcare IoT & PW + SC & $\checkmark$ & $\checkmark$ & $\checkmark$ & $\times$ & $\times$ & $\checkmark$ & $\times$ & WA \\
Protocol 6 \cite{article_26} & Healthcare & PW + SC + BD & $\checkmark$ & $\checkmark$ & $\checkmark$ & $\times$ & $\times$ & $\checkmark$ & $\times$ & WA \\
Protocol 7 \cite{article_55} & Client-Server & TOTP + PUF & $\times$ & $\checkmark$ & $\checkmark$ & $\times$ & $\times$ & $\times$ & $\times$ & WA \\
Protocol 8 \cite{article_56} & Generic IoT & UID + PW + BD & $\times$ & $\times$ & $\times$ & $\times$ & $\times$ & $\times$ & $\times$ & WA \\
Protocol 9 \cite{article_57} & Healthcare & PW + SC + BD & $\checkmark$ & $\checkmark$ & $\times$ & $\times$ & $\times$ & $\checkmark$ & $\times$ & WA \\
Protocol 10 \cite{article_58} & Generic IoT & SSID + SCP & $\times$ & $\times$ & $\times$ & $\times$ & $\times$ & $\times$ & $\times$ & WA \\
\bottomrule
\end{tabular}
}
\smallskip\\
\begin{tabularx}{\linewidth}{X}
\textbf{Description of notations:} LSK: Long-term Shared Key, TGK" Tag Generation Key, HD: Historical Data, HDT: Historical Data Tags, $X^{c}$: factor $X$ is only for Client, $X^{s}$: factor $X$ is only for Server, PW: Password, SC: Smart Card, BD: Biometric Data, TOTP: Time-based Ontime Password, PUF: Physical Unclonable Function, UID: User Identity, SSID: Secret Session Identifier, SCP: Secret Channel-based Parameter, WA: Weak Adversary, SA: Strong Adversary
\end{tabularx}
\end{table*}

\subsection{Comparative performance analysis}

Although most MFA protocols mainly focus on their security guarantee, their performance overhead (or the security-efficiency tradeoff) should not be neglected. In fact, some protocols employ too many authentication factors without considering the penalty in performance. 
However, performance is usually equally critical as security, especially in resource and real-time constrained systems, such as IoT and CPS. Therefore, it is essential to assess the performance of MFA protocols as well. To do such assessment, 
researchers use different types of performance metrics. The most commonly used ones are computation cost (i.e., the amount of resources required to execute the authentication protocol), communication bits (i.e., the amount of data required to be transmitted between entities), communication passes (i.e., the number of messages exchanged between the client and the server to complete the authentication process), authentication time (i.e., the time taken to complete the authentication process), and the storage cost (i.e., the storage size required to complete the authentication process). Therefore, we also used these metrics to perform a comparative performance analysis of the ten protocols investigated. The overall performance analysis is summarized in Table \ref{table_5}. We believe that this helps future researchers to easily obtain the performance information of the protocols in a single reference point. 



\begin{table*}[htb!]
\centering
\caption{Performance comparison of the 10 MFA protocols}
\label{table_5}
\begin{tabularx}{\linewidth}{p{2.9cm} p{3.5cm} p{2cm} p{2cm} p{1.5cm} X X}
\specialrule{.13em}{.05em}{.05em}
Protocol & Computation Cost & Communication \newline Bits & Communication Passes & Time Taken \newline (ms) & Storage \newline Required \\
  \hline
Protocol 1 \cite{article_38} ($\pi_{woFS}$) & $326T_h + 1T_{aed}$ & 3992 & 2 & 22.39 & - \\
Protocol 1 \cite{article_38} ($\pi_{\text{FS}}$) & $328T_h + 4T_{ecc} + 1T_{aed}$ & 4748 & 3 & $115.549^\#$ & - \\
Protocol 2 \cite{article_27} & $18T_h + 14T_x + 2T_{fe} + 2T_{ecc}$ & 1024 & 1 & 198.21 & - \\
Protocol 3 \cite{article_28} & $4T_{me} + (2z + 3)T_m + (2z)T_a + (2z + 26) T_h$ & 2720 & 4 & 11.28 & 3.4GB + 252B \\
Protocol 4 \cite{article_45} & $26T_h + 2T_{ed}$ & 2000 & 3 & - & - \\
Protocol 5 \cite{article_31} & $17T_h\#$ & 1312* & 3 & - & - \\
Protocol 6 \cite{article_26} & $15T_h + 1T_{fe} + 3T_{ed}$ & 2144 & 2 & 8.9385 & - \\
Protocol 7 \cite{article_55} & $4T_h + 2T_p$ & 640 & 1 & - & 404 GB \\
Protocol 8 \cite{article_56} & $5T_h + 2T_{ecc} + 8T_{ed}$ & 1600 & 3 & 80.6 & 480 bits \\
Protocol 9 \cite{article_57} & $1T_{me} + 7T_h$ & 448* & 1 & - & - \\
Protocol 10 \cite{article_58} & $10T_h + 10T_x + 1T_{fe}$ & 896 & 2 & - & - \\
  \toprule
\end{tabularx}
\smallskip
\begin{tabularx}{\linewidth}{X}
\textbf{Description of notations:} $T_h$: hash operation, $T_{\text{ecc}}$: ECC multiplication, $T_{\text{ed}}$: symmetric encryption decryption, $T_{\text{aed}}$: SLHAE encryption and decryption, $T_x$: XOR operation, $T_{\text{fe}}$: fuzzy extraction operation, $T_p$: PUF computation, $T_{\text{me}}$: modular exponential operation, $T_a$: addition operation, $T_m$: multiplication operation, \# : XOR operations ignored, * : Estimated value, -: no information provided.
\end{tabularx}
\end{table*}

\section{Recommended mitigation strategies}\label{mitigation_strategies}

In Section \ref{security_analysis}, we demonstrated that most MFA protocols are not without weaknesses and flaws. Adversaries can potentially target 
such weaknesses to gain unauthorised access 
to systems and applications. In this section, we 
discuss a wide range of 
mitigation strategies 
that can potentially 
address the identified weaknesses and flaws and minimize the risk exposure of the systems and applications. 
Some of the strategies are our suggestions based on our perspectives while some others are best practices that we compiled from existing works. 

\subsection{Mutual authentication}


Mutual authentication can be achieved in many ways. One of the most commonly adopted approaches is using digital certificates in which users authenticate each other through their respective CA-issued public keys. However, due to its high performance overhead, it is not the ideal solution for most resource constrained systems, such as IoT. 
An ideal approach to address this problem is a cryptographic construction where both parties can present provable credentials or factors 
to prove their identity. 
One such approach 
is properly utilizing shared cryptographic keys in a challenge-response manner. For example, the server can compute the response to a random nonce with the shared cryptographic key and compare it with the response computed by the client. If the response matches, the client is authenticated. Similarly, the client will compute the response to another random nonce with the shared cryptographic key and compare it with the response computed by the server. The server is authenticated if the responses match. Kim et al. \cite{article_75} (as demonstrated in Figures \ref{fig:image11} and \ref{fig:image12}) 
used such approach to ensure mutual authentication between entities. It was achieved by verifying messages that were computed using shared secret parameters distributed during the registration phase. Using the shared secret parameter $S_i^1$, the user and the gateway will mutually authenticate each other by verifying the correctness of messages $GM_5$ and $U_i$$M_8$, respectively.

Oh et al. \cite{article_42} also adopted a similar approach (as demonstrated in Figures \ref{fig:image13} and \ref{fig:image14}) 
to achieve mutual authentication. The shared secret keys $K_{MUG}$ and $K_{GSD}$ stored in the Home Gateway (HGW) are used to verify the identity of the user and smart device, respectively. The user and smart device will also utilise the same keys to verify the identity of the HGW to achieve mutual authentication.

Another interesting approach to achieve mutual authentication is by proving a possession of a piece of data that allows both parties to authenticate each other. For example, Zahednejad et al. \cite{article_28} utilized historical data that were shared between the parties, which allows them to prove each other's identity in a challenge-response manner.  

For higher assurance, a more rigorous approach can be adopted using formal methods such as BAN logic \cite{article_68, misc_69} that several researchers used to evaluate the security of their MFA protocols \cite{article_15, article_33, article_34, article_40, article_41, article_42}, to further prove mutual authentication.

\subsection{Distinctiveness of authentication factors}

For an MFA protocol to be fully effective, one of the key design criteria should be 
distinctiveness of its authentication factors. That means, the protocol should involve distinct authentication factors that are drawn from different categories, such as knowledge, possession, inherent, and location factors. 
This is because, the authentication factors of the same category often have a relatively similar security level. Hence, if the adversary manages to break one authentication factor, he can do so on others by applying a similar technique, 
as seen in \cite{article_1}. For example, Jin et al. \cite{article_1} (see the discussion in Section \ref{vuln_proto_1}), used a preshared symmetric key and a tag generation symmetric key as first and second authentication factors for the client, respectively. Here, both are possession factors, and an adversary may apply a similar technique to break them. Therefore, ensuring the distinctiveness of the authentication factors is an essential strategy that MFA protocol designers should follow. 

\subsection{Independence of authentication factors}

The generation of an authentication factor must be unique and independent from others to ensure that the compromise of one authentication factor does not result in the compromise of other authentication factors. For example, an authentication factor must not be derived from 
other authentication factors. In addition, an authentication factor must not be used to protect the confidentiality of another authentication factor. 

By combining multiple independent authentication factors, an adversary must acquire the different authentication factors in order to gain access to the targeted system. 
For example, a three-factor authentication protocol should utilise three independent authentication factors 
where the compromise of one should not affect other. This 
improves the overall security of the MFA protocol and the underlying systems. 
Lee et al. \cite{article_19} demonstrated the independence of authentication factors in its MFA protocol and compromise to any of the authentication factors will not result in further compromise to other authentication factors. Vinoth et al. \cite{article_35} also demonstrated in their MFA protocol that in the event whereby $N-1$ authentication factors are compromised, the remaining authentication factor will stay secret, and the overall authentication process will not be compromised. A snapshot of the approach and authentication steps of Vinoth et al. \cite{article_35} is provided in Figure \ref{fig:image15}. 

\subsection{Key leakage resilience}
Ensuring key leakage resilience is a sophisticated and critical task in the design of an MFA protocol and has to be extensively deliberated to address every possible key leakage scenario. There is no one-size-fits-all design to achieve key leakage resilience and every MFA protocol would require a tailored approach based on the requirements of the systems and the potential threats the system is exposed to. However, there are some guiding principles that can be adhered to achieve the objective. 

Given that the initial authentication process is typically conducted through open communication channels, researchers should ensure that not all parameters utilised in the session key generation are transmitted in clear. In addition, the parameters used in the generation of session keys should not be easily computed using information obtained through the open communication channels and compromise of the authentication factors. One example is the MFA protocol designed by Kwon et al. \cite{article_70}, which is illustrated in Figure \ref{fig:image16}. In this protocol, the equation used to compute the session key is $SK=h(h(N_2||HID_i)||N_3||N_1)$. As shown in Figure \ref{fig:image16}, none of the parameters used in the computation of the session were transmitted in plain through the open communication channel. To obtain the random nonces $N_1$, $N_2$ and $N_3$, the adversary would either need to compromise all three components i.e., User, Gateway, and Sensor Node, or compromise other secret keys or both authentication factors to obtain parameters that are used to protect the random nonces.

Although not specifically discussed in the papers relating to MFA protocols, attaining a high entropy of the cryptographic keys and random nonces, and proper key management is equally important in achieving the objective the MFA protocol aims to deliver. High entropy reduces the likelihood of adversaries from acquiring the sensitive information through 
guessing or brute-force attacks that directly or indirectly results in the compromise of the session keys \cite{misc_72}.
Key security is not only about generation, but comprises of other aspects such as distribution, storage, usage, rotation, and disposal. Therefore, proper key management adhering to guidelines by reputable organisations such as NIST \cite{misc_72, misc_73, misc_74} would provide a higher assurance in the overall security of the authentication process. With technological evolution, cryptographic key lengths once thought to be secure may become vulnerable. Therefore, it is important to keep current with the latest cryptographic standards recommended \cite{misc_76}, and regularly review the security of the MFA protocols.

\subsection{User Anonymity}
Ensuring anonymity can be achieved through the use of anonymous credentials, such as issuing the user with a pseudo identity that will not reveal their actual identity. For example, in the registration phase of the MFA protocol by Kim et al. \cite{article_75} shown in Figure \ref{fig:image11}, the Gateway generates a pseudo identity $TID_i$ using the values of the user’s actual identity $ID_i$, the Gateway’s secret key $K_{GW}$, and a randomly generated nonce $R_{GW}^1$. The pseudo identity $TID_i$ instead of the actual identity $ID_i$ is then used in the authentication process. In addition, the pseudo identity $TID_i$ is refreshed every time with a new pseudo identity $TID_i^{new}$ during the authentication process.
 Alternatively, cryptographic techniques can be employed to protect the anonymity of individuals during the authentication process. These include encryption, hashing, and other mathematical operations such as concatenation, XOR, etc. In the proposed MFA protocol by Bouchaala et al. \cite{article_18}, the actual identity of user $ID_u$ is never transmitted in plain. Instead, cryptographic techniques were employed to generate $PID_u$, which is used in the authentication process, with the following equation: $$PID_u=(ID_u||C_1) \oplus h(B_1||B_3)$$ 
As the value of $B_3$ changes every time, the $PID_u$ value will also be different for every authentication session, making it hard to determine the actual identity of the user.
Similarly, in the design by Lee et al. \cite{article_19}, cryptographic techniques were employed to achieve anonymity. The actual identity $ID_u$, as shown in Figure \ref{fig:image17}, is not transmitted in plain through the open communication channel. Instead, $ID_u$, together with $PWB_i$, were used to derive $UID_u$ and used for the authentication process. $UID_u$ is obfuscated further using cryptographic techniques before it is transmitted to the Cloud Server.

\subsection{Resistance to known attacks}
There are several possible approaches to resist known attacks and actual security mitigation strategies are dependent to specific requirements and constraints present. The following paragraphs will focus on the weaknesses and flaws identified in the protocols analysed in Section \ref{security_analysis} and provide examples on the mitigating strategies that can be applied. 
It is typically not feasible to communicate via encrypted channel during the initial authentication process. The client and server are required to have prior trust established and the required information to establish the secure communication channel which is one of the key objectives of the authentication process. Therefore, appropriate assumption had to be accorded that the MFA protocol is unavoidably susceptible to interception during the initial authentication process, regardless if the adversaries are weak or strong. 
Mitigation strategies discussed in Section 6.3 are applicable in mitigating the potential impacts of client and server impersonation attacks. Researchers have to ensure that only non-sensitive information that does not directly or indirectly affects the security of the MFA protocols is transmitted through the open communication channel. Oh et al. \cite{article_42} demonstrated how impersonation attack is mitigated in their design. Even if the adversary can successfully compromise the first authentication factor i.e. mobile device to obtain the credentials ${A_1,A_2,A_3,A_4, PID_MU}$ and intercept the information i.e., ${PID_{MU}, M_1, C_1, V_{MU}}$ transmitted through the open communication channel, the adversary will not be succeed in impersonating the user as the adversary does not possess information relating to the second authentication factor that is necessary to generate a spoofed authentication request that is valid. 
Similarly, although the adversary can intercept the information ${PID_{MU}, M_3, C_2, V_{MUG}}$ and ${M_5, V_{GSD}}$, the adversary is unable to generate a spoofed response that is valid without information relating to the second authentication factor. In addition, in both scenarios, the adversary is unable to obtain the random nonces i.e.,  $RN_{MU}$ and $RN_G$ which are required in the computation of a valid request and response. 
From the analysis, time-based nonces i.e., timestamps and/or randomly generated nonces were commonly employed to resist replay attacks. However, it is worthy to note that the utilisation of time-based nonces may potentially result in other issue such as DoS attack. The entities involve in the authentication process must have their clocks synchronised in order for authentication process to work effectively. In the event of a significant time difference or clock drift between the entities, it can lead to authentication failures, resulting in DoS. It is therefore recommended to adopt the use of randomly generated nonces to eliminate the need to depend on external factors, in this case time synchronisation. As demonstrated in \cite{article_42} and shown in Figure \ref{fig:image13}, randomly generated nonces are employed to ensure the “freshness” of the authentication request and is able to resist replay attacks. However, if time-based nonces must be adopted for any reasons, additional control measures shall be implemented to ensure accuracy in time synchronisation to minimize the occurrence of time discrepancies.

\subsection{Adversary assumption}
Assumption of the adversary’s capability is vital in determining the necessary security mitigation strategies required to satisfy the security requirements and reduce the likelihood of the risks occurring to the lowest possible. As indicated by Zahednejad et al. \cite{article_28}, assuming a weak adversary that is only capable of obtaining a small portion of information may not truly reflect the actual reality. It is therefore important to adopt a more realistic approach and accord the adversary with the necessary respect so that security mitigation strategies can designed and implemented to successfully thwart the attacks.

\subsection{Summary}
Table \ref{table_mitigation_strategies} provides a summary of the possible mitigation strategies that can be applied to mitigate the weaknesses and flaws identified in Section \ref{security_analysis}. As shown, the mitigation strategies are nothing out of the ordinary and have been implemented in the design of MFA protocols by other researchers \cite{article_18, article_19, article_28, article_35, article_42, article_70, article_75}.

\begin{table*}[htb!]
\caption{Summary of possible mitigation strategies}
\label{table_mitigation_strategies}
\begin{tabularx}{\linewidth}{|p{3.5cm}|X|}
\toprule
Security Evaluation Criteria & Possible Mitigation Strategies \\
\midrule
Mutual Authentication & 
-- Both parties presenting appropriately provable credentials or factors to each other 
to prove their identity.\newline
-- Cryptographic computations and verifications using shared secret keys.\newline
-- Perform security analysis using methods such as BAN logic to formally prove the security properties.\\ \hline
Distinctiveness of factors & 
-- Ensure that the authentication factors are from distinct categories.\\ \hline
Independence of factors & 
-- Generation of authentication factor must be unique and cannot be derived from the knowledge of another authentication factor.\newline
-- One authentication factor must not be used to protect the confidentiality of another authentication factor.\\ \hline
Key Leak Resilience & 
-- Do not transmit all parameters utilized in the session key generation in clear.\newline
-- Parameters utilized in the generation of session keys should not be easily derived from the information transmitted in clear.\newline
-- Achieve high entropy of the cryptographic keys and random nonces.\newline
-- Perform proper key management in line with guidelines by reputable organizations such as NIST.\\ \hline
Perfect-forward secrecy & 
-- Construct the MFA protocol (including its authentication factors) in such a way that ensures the leakage of long-term keys of both client and server cannot compromise the security of previous sessions. In other words, previous session keys should not be computed from long-term keys of the client and the server. \\ \hline
User Anonymity & 
-- Employ strong user privacy-preserving approaches (even beyond the use of pseudonymous identities).\newline
-- Utilize cryptographic techniques to protect the anonymity of individuals.\\ \hline
Resistance to Known Attacks i.e., impersonation and DoS attacks & 
-- Do not transmit sensitive information in clear that directly or indirectly affects the security of MFA protocols.\newline
-- Protect parameters such as random nonces using cryptographic techniques and ensure that compromise to $N-1$ authentication factors would not affect the security of the MFA protocols. \\ \hline
Adversary Assumption & 
-- Adopt a realistic approach and accord the adversary with appropriate capabilities.\\
\bottomrule
\end{tabularx}
\end{table*}
\section{Conclusion}\label{conclusion}

With the focus on authentication, several MFA protocols specifically tailored for the various domains have been developed in recent years. However, common but yet critical security criteria that should have been considered and applied were observed to be omitted in some of the proposed MFA protocols. In some cases, claims of the MFA protocols were capable of satisfying certain security criteria were proved otherwise. In this work, we reviewed several MFA protocols and analyzed potential vulnerabilities heuristically. In particular, we systematically analyzed security flaws in the construction of the protocols using a set of security evaluation criteria we employed. 

Consequently, we managed to identify several vulnerabilities in ten of the MFA protocols. We provided a detailed discussion of the vulnerabilities identified. We also highlighted relevant mitigation strategies for those vulnerabilities. We believe that the consolidated information would provide a single reference point for researchers to be aware of the potential security issues that require attention and apply the necessary mitigation strategies when designing MFA protocols. It is also worth noting the importance of performance alongside its security. The performance of the MFA protocol should not be overlooked. Complementing it with an efficient performance will improve the adoption rate of the secure MFA protocol, thus enhancing the overall security.

To further strengthen the design and implementation of an MFA protocol, a security-by-design approach \cite{misc_78} should be considered. In the future, this work can be further extended by employing a formal analysis of the protocols. Emerging security concerns, such as the security risk posed by the advent of quantum computing, can also be included as additional evaluation criteria.  

\section*{Acknowledgment}

This research is supported by the National Research Foundation, Singapore and Infocomm Media Development Authority under its Trust Tech Funding Initiative (DTC-T2FI-CFP-0002). Any opinions, findings and conclusions or recommendations expressed in this material are those of the author(s) and do not reflect the views of National Research Foundation, Singapore and Infocomm Media Development Authority.

\bibliographystyle{ACM-Reference-Format}
\bibliography{10_references}
\appendix
\counterwithin{figure}{section}
\counterwithin{table}{section}

\onecolumn
\section{Supplementary Figures}

\begin{figure*}[htb]
    \centering
    \includegraphics[scale=0.35]{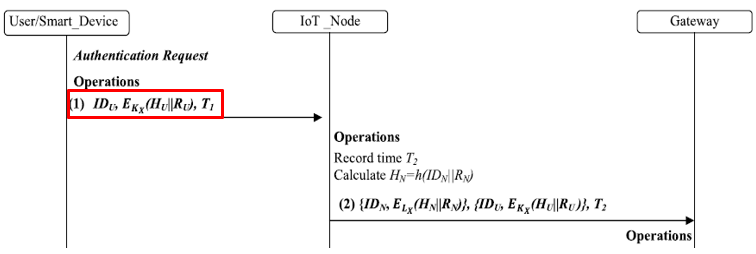}
    \caption{Snapshot of the MFA protocol in \cite{article_56} where $ID_u$ is transmitted in plain}
    \label{fig:image9}
\end{figure*}

\begin{figure*}[htb]
    \centering
    \includegraphics[scale=0.4]{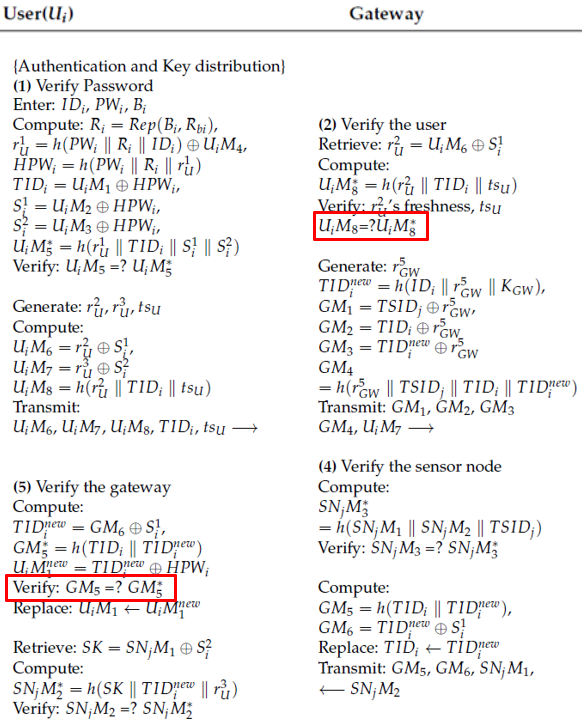}
    \caption{Snapshot of the checks performed to mutually authenticate user and Gateway in \cite{article_75}}
    \label{fig:image12}
\end{figure*}

\begin{figure*}[htb]
    \centering
    \includegraphics[scale=0.45]{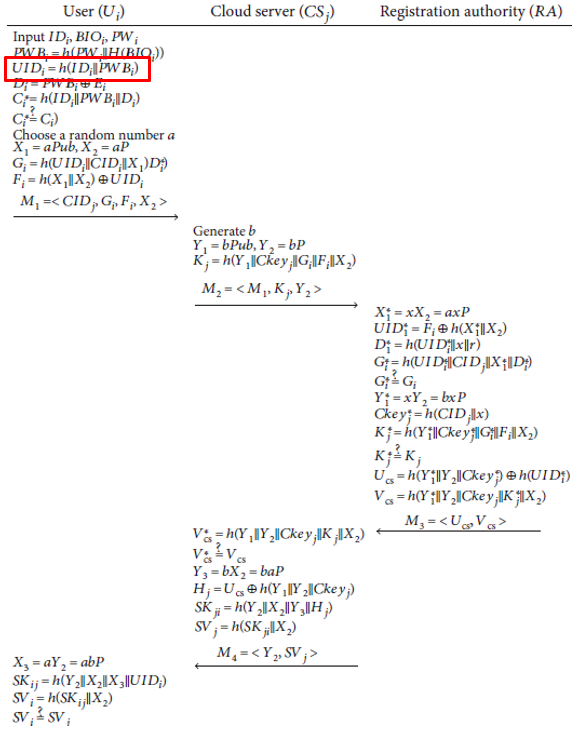}
    \caption{Snapshot of the authentication phase in \cite{article_19} whereby $UID_u$ is used to protect the actual identity}
    \label{fig:image17}  
\end{figure*}

\begin{figure*}[htb]
    \centering
    \includegraphics[scale=0.4]{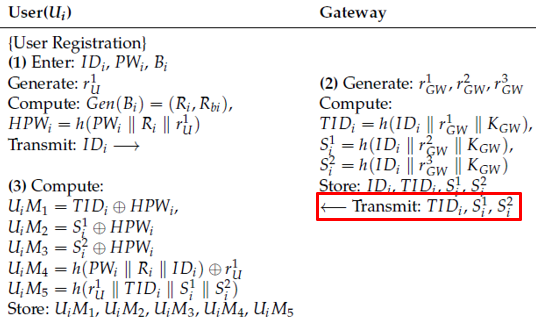}
    \caption{Snapshot on the distribution of shared secret parameter $S_i^1$ in \cite{article_75}}
    \label{fig:image11}
\end{figure*}

\begin{figure*}[htb]
    \centering
    \includegraphics[scale=0.44]{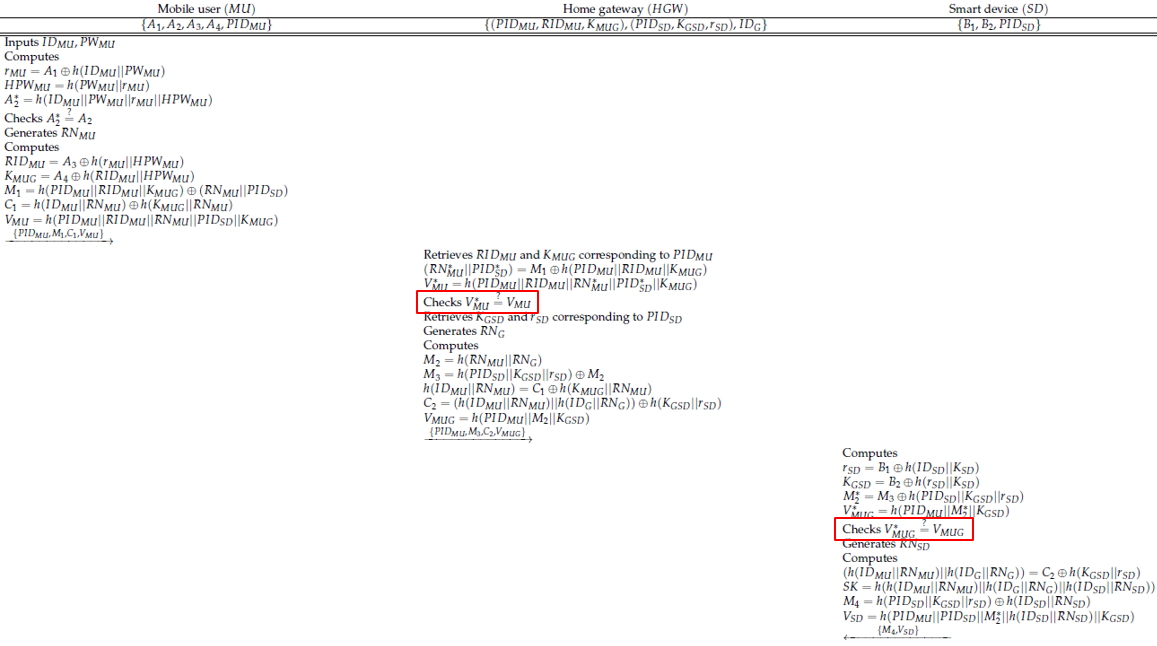}
    \caption{Snapshot of the checks performed to mutually authenticate user, HGW and smart device in \cite{article_42}(part 1)}
    \label{fig:image13}
\end{figure*}

\begin{figure*}[htb]
    \centering
    \includegraphics[scale=0.29]{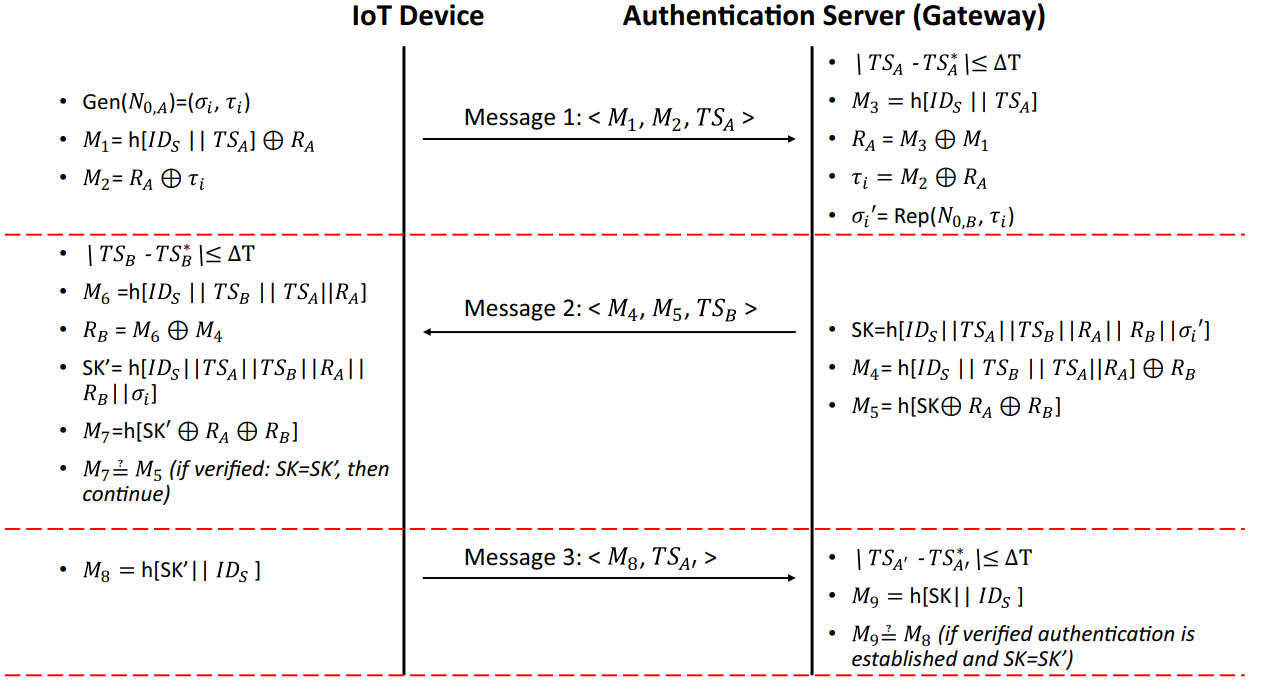}
    \caption{Snapshot of the authentication and key agreement phase in \cite{article_56}}
    \label{fig_protocol10_auth}  
\end{figure*}

\begin{figure*}[htb]
    \centering
    \includegraphics[scale=0.6]{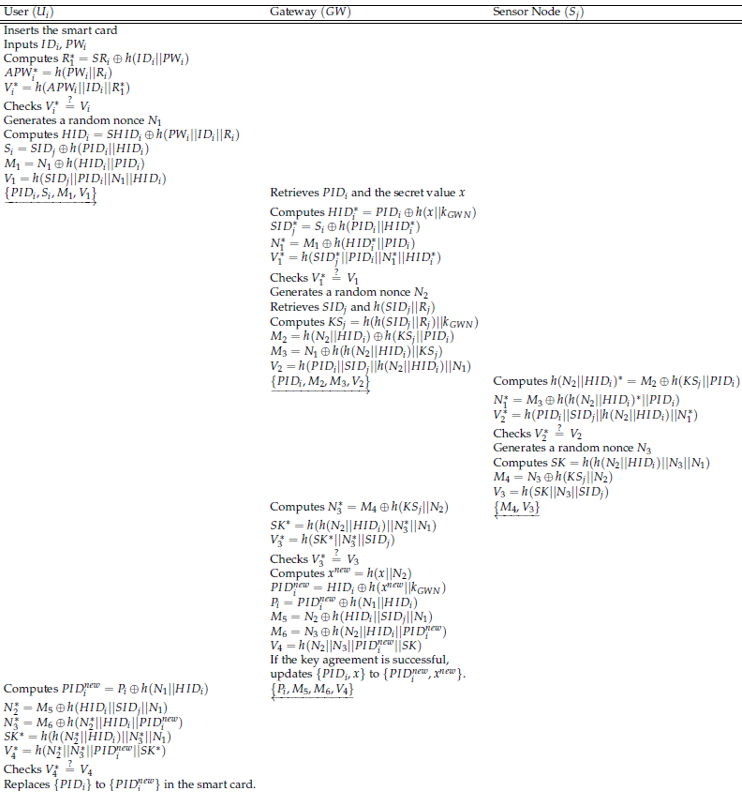}
    \caption{Snapshot of the authentication phase demonstrating key leakage resiliency in \cite{article_70}}
    \label{fig:image16}
\end{figure*}

\begin{figure*}[htb]
    \centering
    \includegraphics[scale=0.44]{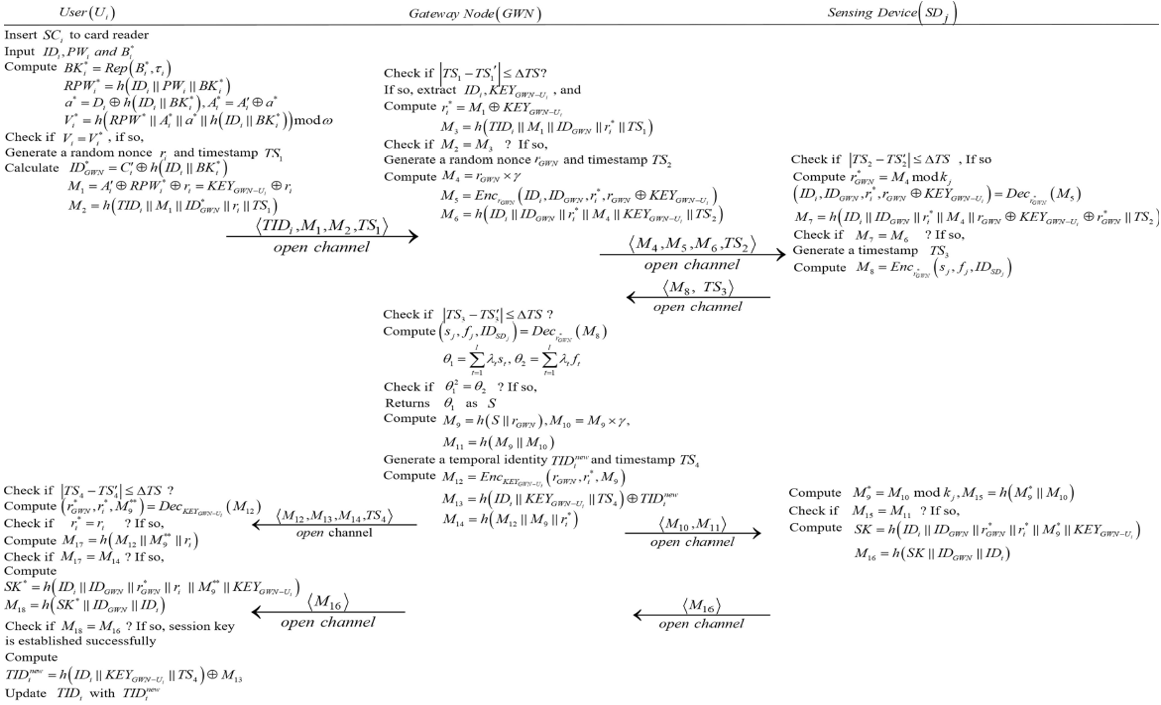}
    \caption{Snapshot of the authentication phase in \cite{article_35} showing authentication factors’ independence}
    \label{fig:image15}
\end{figure*}

\begin{figure*}[htb]
    \centering
    \includegraphics[scale=0.49]{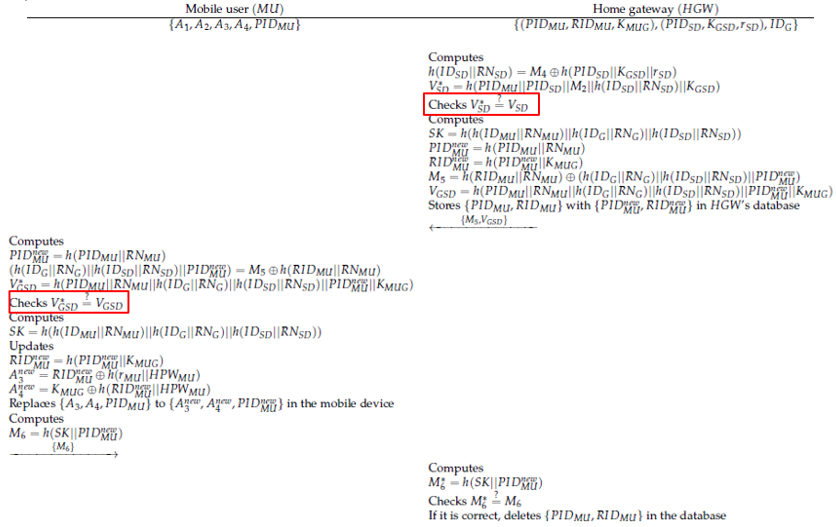}
    \caption{Snapshot of the checks performed to mutually authenticate user, HGW and smart device in \cite{article_42}(part 2)}
    \label{fig:image14}
\end{figure*}


\end{document}